\documentclass[twocolumn]{aastex62}

\graphicspath{{./}{figures/}}

\accepted{\today}
\submitjournal{ApJ}

\shorttitle{Chemical Evolution in Protoplanetary Disks with Dust Substructures}
\shortauthors{Alarc\'on et al.}
\begin{document}

\title{Chemical Evolution in a Protoplanetary Disk within Planet Carved Gaps and Dust Rings}

\correspondingauthor{Felipe Alarc\'on}
\email{falarcon@umich.edu}

\author[0000-0002-2692-7862]{Felipe Alarc\'on}
%\altaffiliation{Editor-in-Chief}
\affiliation{Department of Astronomy, University of Michigan,
323 West Hall, 1085 S. University Avenue,
Ann Arbor, MI 48109, USA}

%% The \author command can take an optional ORCID.
\author[0000-0003-1534-5186]{R. Teague}
\affiliation{Department of Astronomy, University of Michigan,
323 West Hall, 1085 S. University Avenue,
Ann Arbor, MI 48109, USA}
\affiliation{Center for Astrophysics | Harvard \& Smithsonian, 60 Garden Street, Cambridge, MA 02138, USA}

\author[0000-0002-0661-7517]{Ke Zhang}
\altaffiliation{NASA Hubble Fellow}
\affiliation{Department of Astronomy, University of Wisconsin-Madison, 
475 N Charter St, Madison, WI 53706}
\affiliation{Department of Astronomy, University of Michigan, 
323 West Hall, 1085 S. University Avenue, 
Ann Arbor, MI 48109, USA}

\author[0000-0003-4179-6394]{E. A. Bergin}
\affiliation{Department of Astronomy, University of Michigan,
323 West Hall, 1085 S. University Avenue,
Ann Arbor, MI 48109, USA}

\author{M. Barraza-Alfaro}
\affiliation{Max Planck Institute for Astronomy, Königstuhl 17, D-69117 Heidelberg, Germany}


\begin{abstract}

Recent surveys of protoplanetary disks show that  substructure in dust thermal continuum emission maps is common in protoplanetary disks.  These substructures, most prominently rings and gaps, shape and change the chemical and physical conditions of the disk, along with the dust size distributions.
In this work, we use a thermochemical code to focus on the chemical evolution that is occurring within the gas-depleted gap and the dust-rich ring often observed behind it. The composition of these spatial locations are of great import, as the gas and ice-coated grains will end up being part of the atmospheres of gas giants and/or the seeds of rocky planets. Our models show that the dust temperature at the midplane of the gap increases, enough to produce local sublimation of key volatiles and pushing the molecular layer closer to the midplane, while it decreases in the dust-rich ring, causing a higher volatile deposition onto the dust grain surfaces. Further, the ring itself presents a freeze-out trap for volatiles in local flows powered by forming planets, becoming a site of localized volatile enhancement.  Within the gas depleted gap, the line emission depends on several different parameters, such as: the depth of the gap in surface density, the location of the dust substructure, and the abundance of common gas tracers, such as CO. In order to break this uncertainty between abundance and surface density, other methods such as disk kinematics, become necessary to constrain the disk structure and its chemical evolution. 
\end{abstract}


%% Observe the use of theLaTeX \label

\section{Introduction}\label{sec:intro}

Planets are born within young disks that formed from the collapse of a centrally concentrated dense core within molecular clouds. It is during the gas-rich protoplanetary disk stage that gas giant planets capture the volatile-rich gaseous material that will be part of their atmospheres. Therefore, giant planets have to be formed before the gas dissipates, a process that takes a few Myr \citep{Alexander_2014}. Further, the physical state of protoplanetary disks  is constantly evolving \citep{Armitage..2011,Williams..&..Cieza..2011,Andrews_2020}, and these changes can affect how planets form. For example, the initial composition and elemental budget  of a planet would be different depending on the time and location of formation \citep{Oberg..MurrayClay..Bergin..2011,Cridland..et..al..2016, Alessi..et..al..2017}.

%Understanding the interactions between the gas chemistry and the solid state chemistry is also important for the  terrestrial worlds as  the carriers of life-bearing elements are highly volatile  (C, H, O, N, and S) and are thus difficult to incorporate in rocks in close proximity to young stars.  
%This does not only depend on the planet's distance from the host star \citep{Kaltenegger..2017}, but also on the parent disk structure. 

Recent ALMA surveys of protoplanetary disks have shown that disk substructures are ubiquitous in protoplanetary disks, particularly axisymmetric ones \citep{Zhang..et..al..2016,Long..et..al..2018,Andrews..et..al..2018,Huang..et..al..2018, Zhang..et..al..2018}. Among these axisymmetric substructures, the most common features are gas-depleted gaps \footnote{In this paper we will refer to depletion as any decrease of either gas (H$_2$ mass) or dust surface density from a smooth profile,  while the deposition of volatiles into grains will be referred as freeze-out.} and dust-rich rings. %\textcolor{red}{Nevertheless, disk substructures are not the only source of chemical changes in a protoplanetary disk}. There are other factors with the potential to change the chemistry, such as: the luminosity and spectrum of central star, the high-energy X-ray and ultraviolet (UV) radiation fields, and the disk's evolving physical properties.
 Within that context, it becomes important to constrain the chemical evolution during the early stages of planet formation. Previous works have explored the chemical changes taking place during disk evolution. Just to give a few examples: \citet{Eistrup..et..al..2016,Eistrup..et..al..2018} consider different initial abundances and the chemical evolution of the midplane including planet accretion; \citet{Facchini..et..al..2017,Booth_et_al_2017,Krijt_et_al_2018} take into account dust evolution. \cite{Cleeves..et..al..2013} explore the question of cosmic-ray ionization in protoplanetary disks, while significant emphasis has been places on the evolution of CO abundance \citep{Reboussin..et..al,Yu..et..al..2017,Kamber..et..al..COI,Kamber..et..all..COII,Bosman..et..al..2018}.
 More recently, \citet{Facchini..et..al..2018} post-process hydrodynamic simulations within a  thermochemical model, making predictions of line emission inside  planetary gaps and \cite{Vandermarel..et..al..2018} look at the observable effects that different gap-carving processes have. Overall, the consensus is clear: gaps and rings will set the local conditions around forming planets.
 %and   may alter the elemental partitioning of material that is provided to them. 

 %Moreover, by knowing the chemical evolution in a protoplanetary disk and a planet's composition it could even be possible to trace back its formation location. The understanding of the chemical evolution in a disk will not only allow to \textcolor{red}{explore} the possible habitability of a planet, but also understand the possible migratory motions that a planet has gone through its early stages. 

There are several possible mechanisms to explain substructure formation. The primary theory is that young planets carve gaps in the dust and gas \citep{Lin_1986}, inducing pressure bumps within the edges of the gaps \citep{Armitage..2010, Pinilla..et..al..2012, Pinilla..et..al..2015,Zhang..et..al..2018}.  The outer pressure bump then serves as a location where drifting dust will pile up, potentially evolve, and be observed as  a ring \citep{Dullemond..et..al..2018, Andrews..et..al..2018}. However, there are  other possible scenarios that produce pressure bumps. One of those scenarios takes place in in the outer edge of the MRI dead zone in MHD simulations \citep{Flock..et..al..2015}; another one occurs in the discontinuities produced by ice lines of abundant molecules, inducing dust growth \citep{Zhang..et..al..2016, Pinilla..et..al..2017}. Both, gaps and rings, are explained by the same mechanisms and are ubiquitous in protoplanetary disks. 

\begin{figure*}
%\plotone{Density.pdf}
\includegraphics[width=1.\textwidth]{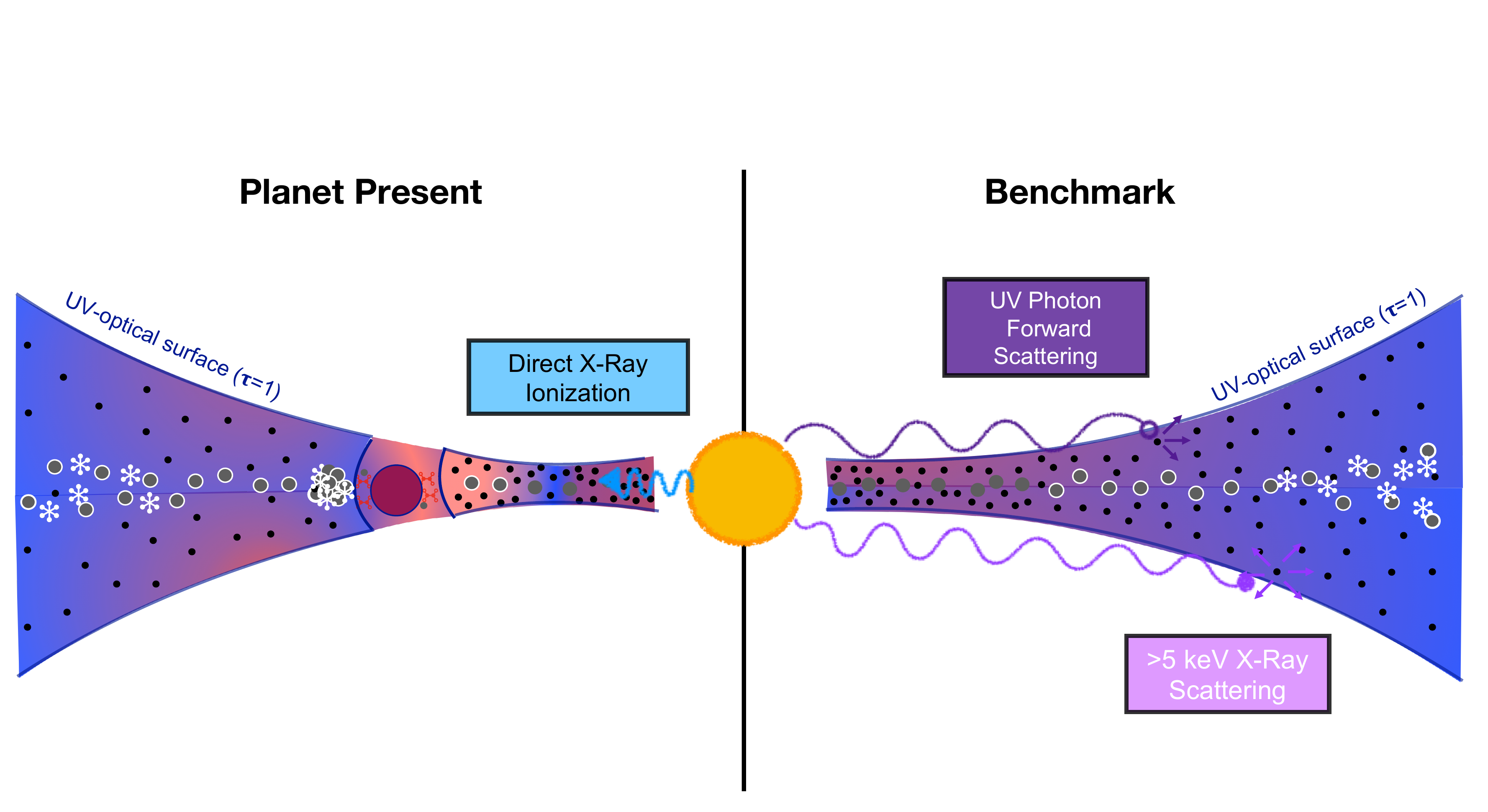}
\caption{Sketch illustrating the main effect expected with the presence of a gap-ring structure produced by a planet. High energy scattering and ionization occurs in both cases, but how they will affect the disk changes. The ionization fraction inside the gap should increase due to lower opacity to UV photons. In the ring, there is freeze-out of molecules due to lower temperatures and also the larger dust surface area triggers a faster dust grain chemistry. \label{Fig:sketch}}
\end{figure*}

The gap-ring substructure changes the local dust surface density and because of that, the dust opacity. Depending on the location of the gap-ring substructure, variations in the dust surface density could change the chemistry in major ways \citep{Cridland..et..al..2017}.
Figure \ref{Fig:sketch} is a sketch of the phenomenology we expect to take place inside the gap-ring substructure. For example, there is a pile-up of dust grains in the ring and the gap gets hotter closer to the midplane because its lower optical depth. Further,the propagation of high energy photons, both FUV and X-rays, is enhanced, altering the local chemical equilibrium \citep{Walsh..et..al..2012}.  The high energy photons  photodissociate gas-phase species or photodesorb icy grain mantles, potentially fostering unique chemical signatures that would not otherwise be present \citep{Cecchi..&..Aiello..1992,Zhu..et..al..2014,Oberg..et..al..2009I, Walsh..et..al..2010,Walsh..et..al..2012,Bruderer..2013}. The net surface area of dust grains within the rings is also much larger, which presumably fosters further volatile deposition and growth.  

%Given that reactions on grain surfaces are posited as central to organic formation \citep{Tielens..Hagen..1982,van_dishoeck_2014}, this may have important implications for development of chemical complexity.

%Understanding the chemistry in these dust-rich substructures could give important insights on the composition of rocky planetary building blocks as well. These rocky blocks are the initial stage for the cores of giant planets or the seeds that create a rocky planet, like Earth. Given the amount of dusty material present in rings, they could be an ideal place for rocky planet formation. Furthermore, \cite{Dullemond..et..al..2018} reported that narrow rings in the DSHARP survey could have an amount of solids of the order of ten Earth masses, which allow them to trigger the streaming instability \citep{Youdin..&..Goodman..2005}, enabling the build-up of planetesimals there. Giant planets carve gaps in their host disk \citep{Lin_1986}, by knowing the chemical composition inside the gap we could constrain  some physical parameters of the protoplanet's atmosphere \citep{Seager..&..Deming..2014}, linking it with the exoplanet population discovered so far.

In this work, our goal is to understand the effect of a gap-ring   substructure in the temperature and chemistry of a protoplanetary disk. We focus on the changes in the physical conditions inside the gap and the ring, and how these changes lead to the given abundance and distribution of important chemical tracers.

This paper is organized as follows: Section \ref{sec:model} explains the setups of our thermo-chemical models, including the assumptions and testing goals. Section  \ref{sec:results} shows the main results of the simulations; the chemical properties and the synthetic ALMA images for a chosen set of molecular transitions. In Section 4, we discuss the possible implications of our results. Finally, we present a brief summary of this work in Section \ref{sec:Summ}.

\section{Modeling and Methodology} \label{sec:model}

We study the chemistry and radiation fields in protoplanetary disks with a dust substructure, using the 2D thermo-chemical code \texttt{rac2d} \citep{Fujun..&..Bergin..2014}. The code calculates the dust thermal equilibrium with a Monte Carlo algorithm. After the thermal equilibrium is attained, the code iterates using chemical kinetics to reach a quasi-chemical equilibrium. 

Our chemical network includes 484 species with 5830 chemical reactions. The gas reaction rates in the chemical network are mainly taken from the UMIST 2006 database \citep{Woodall..et..al..2006}. Furthermore, the code includes reactions on the dust surface, which are taken from \cite{Hasegawa..et..al..1992}; the photodesorption of H$_2$O and OH by Ly-$\alpha$ photons used yields from \citet{Oberg..et..al..2009II}; full Ly-$\alpha$ radiation transfer and inclusion into photodissociation rates, and the adsorption/desorption of molecules on dust grains using formalism as described in \citet{Fujun..&..Bergin..2014}, which contains a thorough explanation of the code's functioning. 

The time evolution in \texttt{rac2d} is solved through logarithmic time steps in isolated cells. Our simulations do not take into account any turbulent mixing or radial motion in the gas or dust. During the chemical iterations, the code updates the heating/cooling sources in the gas, which allows us to find the gas temperature decoupled from the dust. The decoupling between gas and dust temperature becomes significant in the upper layers of the disk \citep{Kamp..vanZ..2001,Jonkheid..et..al..2004}, where is particularly important to make accurate predictions of the chemical rates and the molecular line emission in synthetic images.

Our goal is to analyze the effect of the gaps and rings on the chemical evolution of the disk. In order to do that, we include a static substructure in the density structure from the beginning of the simulations. Then, we observe how the substructure changes the final composition of the disk when compared with a disk model without them. The 2D nature of the code lets us analyze the vertical structure and the warm molecular layer as well. 

Once we have run the simulations for 1 Myr, we compare our models, their thermal fields and the distribution of molecules in the network, having a particular interest on the chemistry inside the disk substructure. Finally, we create synthetic ALMA images to predict how the local substructure within gas and dust alters the expected radial emission profiles.

\subsection{Simulation Setup}

Stellar irradiation is the main heating source of the disk and along with the dust dustribution they set the thermal profile of the disk. Likewise, UV and X-rays photons dominate the chemistry in the upper layers through molecular ionization and photodissociation. 

Given that photodissociating photons play a considerable role in the disk chemistry, we use a large number of photon packages for the X-rays and the UV frequencies to minimize the photon noise in the Monte-Carlo simulation. We run the simulations using $5\times 10^6$ photons packages for the optical and longer wavelengths, but we include a refinement factor of 10$^{3}$ for Ly~$\alpha$ photons, 5$\cdot \ 10^{3}$ for UV photons and 10$^{6}$for X-rays. By using the refinement factor, we increase the number of photons packages in the respective spectral window by the corresponding refinement value. Even using a refinement for the UV photons, the dust opacity does not let the UV photons to reach the midplane. Nevertheless, the UV flux reaching the midplane and its associated photodissociation rate are negligible compared to the optical irradiation and the gas-phase chemistry, so it does not change our final results. Further \cite{Agundez_2018}  show that FUV photodissociation rates for T Tauri disks is negligible for $z/r<0.3$ at 100 au.

In our basic or benchmark model, the surface density, $\Sigma_{\rm basic}$, follows the self-similar solution from \cite{Lynden-Bell..Pringe..1974}, i.e.,

\begin{equation}\label{eq: Self-Similar}
    \Sigma_{\rm basic}(r) = \Sigma_0 \Big(\frac{r}{r_0} \Big)^{-\gamma}\exp \Big[-\Big(\frac{r}{r_e} \Big)^{2-\gamma}\Big],
\end{equation}

\noindent where $r_e$ is the radius of the exponential cut-off, $r_0$ is the characteristic radius for the disk and $\Sigma_0$ is the surface density of the disk at $r=r_0$. In our basic model, the gas and dust components have the same profiles, where $\Sigma_0$ is the only changing value.
For the disk with the dust substructure, we add the ring and the gap as modulations of the self-similar solution. Those modulations are further discussed in Sect. \ref{SSect: DiskSu}.  
We aim to include how the gas and the distinct dust populations alter the chemistry and physical conditions in the disk. So, we set different vertical distributions for them, but with the same functional form. 
Both gas and dust have vertical gaussian distributions \citep{Armitage..2010}:

\begin{equation}\label{eq: vertical Dist}
\rho(r,z) = \rho_0(r)\exp\Big(-\frac{z^2}{2h^2} \Big),
\end{equation}

\noindent where $\rho_0 = \frac{\Sigma_{\mathrm{basic}}(r)}{\sqrt{2\pi}h}$ and $h$ is the scale height, which we vary for each population. Particularly for the gas, the aspect ratio is set to be $h/r = 0.08$ at $r=100$ au.  The scale height also has a radial dependence through a power-law with a flaring index $\psi$:

\begin{equation}\label{eq: Flaring}
h(r) = h_{c}\Big(\frac{r}{r_{c}} \Big)^{\psi}.
\end{equation}

\noindent The flaring index is a common value for the gas and dust populations. We set it with a value of $\psi$=1.22. 

Besides the gas, we include two main dust populations differentiated by the grain size they represent and an additional third component only within the inner 3.5 au. The two main populations are composed of a mixture of 80\% silicates and 20\% graphite. The optical opacities where taken from \cite{Draine..&..Lee..1984}, while the X-rays opacities come from the prescription in \cite{Bethell..Bergin..2011}. Both dust populations follow a grain size distribution with a standard power-law, $n(a) \propto a^{-3.5}$, which is taken from \cite{Mathis..et..al..1977}. We also set a differentiated vertical distribution for different dust populations to consider the dust settling effect \citep{Dullemond..&..Dominik..2004}.

The first dust population is composed only of small grains (0.005 $\mu$m$<a_{\rm grain}<$ 1 $\mu$m). Although the small grains populations contains only 12.5\% of the total dust mass, it has a considerable share of the total dust surface area in the disk, where most of the dust-gas reactions happen and effectively governs the gas/dust interactions, except in the ring midplane. Given that the small grains are more coupled to the gas, we set an aspect ratio of $h/r = 0.08$ at $r$=100 au for them.

The second dust population contains the large grains and the majority of the dust mass (87.5 \%). This population has a maximum grain size $a_{\rm max}=1$ mm and a minimum size  $a_{\rm min}$=0.005 $\mu$m.  The dust settling was implemented as follows: the large grains, which are more concentrated in the midplane, have an aspect ratio $h/r = 0.02$ at $r=100$ au \citep{Boehler..et..al..2013}. The settling becomes important for large grains because it changes the transparency to UV photons producing a larger UV-Photon Dominated Region. A UV-Photon Dominated Region, along with the gap, could enhance the emission of small hydrocarbons \citep{Bergin..et..al..2016}. 

The third population has the same size distribution than the first one, but it is only composed of silicate grains in the inner 3.5 au. We are mostly focus in the gap at 100 au, but there is a transition in the dust composition inside and outside the water snow line, so we included the third population just for completeness.

The disk was modeled using a logarithmic grid in cylindrical coordinates $(r,z)$, with 300 cells in the radial direction between 0.1 and 300 au.

\begin{deluxetable}{c|c}
\tablecaption{Simulation Setup\tablenotemark{a} \label{tab:pars table}}
\tablehead{
\colhead{Parameter} & \colhead{Value} 
}
\startdata
Disk Mass  & 70 $M_{\rm Jup}$ \\
Dust Mass  & 0.8 $M_{\rm Jup}$ \\
Stellar Radius & 2.4 $R_{\odot}$ \\
Stellar Mass & 0.9  $M_{\odot}$\\
Stellar Temperature  & 4250 K\\
UV Luminosity & 2.91$\cdot 10^{32}$ erg s$^{-1}$ \\
X-Rays luminosity & 3.69$\cdot 10^{30}$ erg s$^{-1}$ \\
T$_{\rm X-Rays}$  & 10$^7$ K \\
$\zeta$ ionization rate & 1.36$\cdot 10^{-17}$ s$^{-1}$ \\
$t_{\rm disk}$ & 1 Myr \\
$\gamma$ & 1 \\
$\psi$ & 1.22  \\ 
$r_c$ & 100 au\\
r$_{\rm gap}$ & 100 au \\
$\delta_{\rm gap,large dust}$ &  0.0025\\
$\delta_{\rm gap,gas}$ &  0.085\\
w$_{\rm gap}$ & 16 au \\
r$_{\rm ring}$ & 120 au \\
$\delta_{\rm ring}$ & 26 \\
w$_{\rm ring}$ & 4.11 au\\
\enddata
\end{deluxetable}

\subsubsection{Disk Substructure}\label{SSect: DiskSu}

We include the gap-ring substructure by gaussian modulations of the basic surface density profiles, increasing or decreasing locally the amount of material. We use gaussian profiles because, rather than step functions, they produce smooth transitions between the gaps/rings and the parametric surface density profiles of the disks. Despite that the sharpness of the feature will be dictated by the process forming it, the current spatial resolution precludes moving beyond a smooth gaussian prescription.

\begin{enumerate}
    \item {Gap}:
  We set the gap as a region where there is a local depletion of material for both dust populations and the gas. We characterize them by their location $r_{\rm gap}$, the depth of the gap $\delta_{\mathrm{gap}}$, and a width $w_{\mathrm{ gap}}$. In this case, we modulate the basic surface density using a gaussian function:
    \begin{equation}\label{eq:gap}
    \Sigma(r) = \Sigma_{\rm basic}(1 - (1 - \delta_{\rm gap})\exp(-(r-r_{\rm gap})^2/2w_{\rm gap}^2)).
    \end{equation}
    
    \noindent Thus, The location, $r_{\rm gap}$ , sets the center of the material depletion; the depth, $\delta_{\mathrm{gap}}$, represents the ratio between the surface density in the center of the gap with respect to the undepleted case; and the width, $w_{\mathrm{ gap}}$, is the standard deviation of our gaussian gap. It is noteworthy that instead of being absolute, the depth of the gap is relative to the surface density at that location in an undepleted case.
    
\item{Ring}:
The ring is parametrised in a similar way to the gap, but the main difference is that it is an enhancement of material and it only applies to large grains (mm-sized grains). We only apply it to large grains because dust trapping is more efficient for larger sizes \citep{Birnstiel_et_al_2010,Pinilla..et..al..2012, Weber..et..al2018}. Further, high dust density enhances dust growth increasing the amount of large grains, while small grains are more coupled to the gas. By being more coupled to the gas, the small grains feel a stronger influence of the background gas pressure profile and thus do not form rings as readily as larger, less well coupled grains do. \cite{Dullemond..et..al..2018} showed that at current observational resolution, dust rings can be modeled with gaussian functions, so we use gaussian modulation of the dust surface density to characterize the ring.
    Rings are characterized by their location $r_{\rm ring}$, an amplitude $\delta_{\rm ring}$, and a width $w_{\rm ring}$. The final expression for the surface density including a ring is:
    
    \begin{equation}\label{eq:ring}
            \Sigma(r) = \Sigma_{\rm dust}(1 + \delta_{\rm ring}\exp(-(r-r_{\rm ring})^2/2w_{\rm ring}^2))
    \end{equation}

\end{enumerate}

A final summary of the parameters in our simulations can be found in Table  \ref{tab:pars table}. We based the main gap-ring substructure of our simulated disk on AS 209, which, besides having an axisymmetric dust emission with a large gap, has been previously modeled with a giant planet carving the gap \citep{Salyk_et_al_2013,Banzatti_et_al_2017,Fedele..et..al..2018,Avenhaus_et_al_2018,Zhang..et..al..2018, Guzman..et..al..2018}.  However, even though the actual parameters of the  substructure will depend on the conditions of each particular disk and the planet creating the substructure, its functional form is rather generic \citep{Zhang..et..al..2018,Dong..&..Fung..2017}. Despite that we based our model in AS209, instead of making a source specific model, we explore the effect of the substructure on the thermo-chemical structure of the disk. Small grains and gas have a less depleted substructure than large grains as they couple to the gas. Large grains are not dynamically coupled to the gas so they get trapped in the pressure maxima at the gap's edge. The aim is to find the physical conditions produced by such substructure on the disk and the chemical networks that it could trigger. Thus, we use different depletion factor for small and large grains. We used a depletion for large grains similar to the one used by \cite{Facchini..et..al..2017}, while our gas and small grains depletion is similar to the value used by \cite{Favre..et..al..2019} (see Table \ref{tab:pars table}). 

%% Salyk: Radius, Temp, Solar Mass
%% Banzatti: Accretion lum
%% Avenhaus: Dust mass, flaring
%% Andrews: Mass of star

\begin{figure}
\includegraphics[width=.47\textwidth]{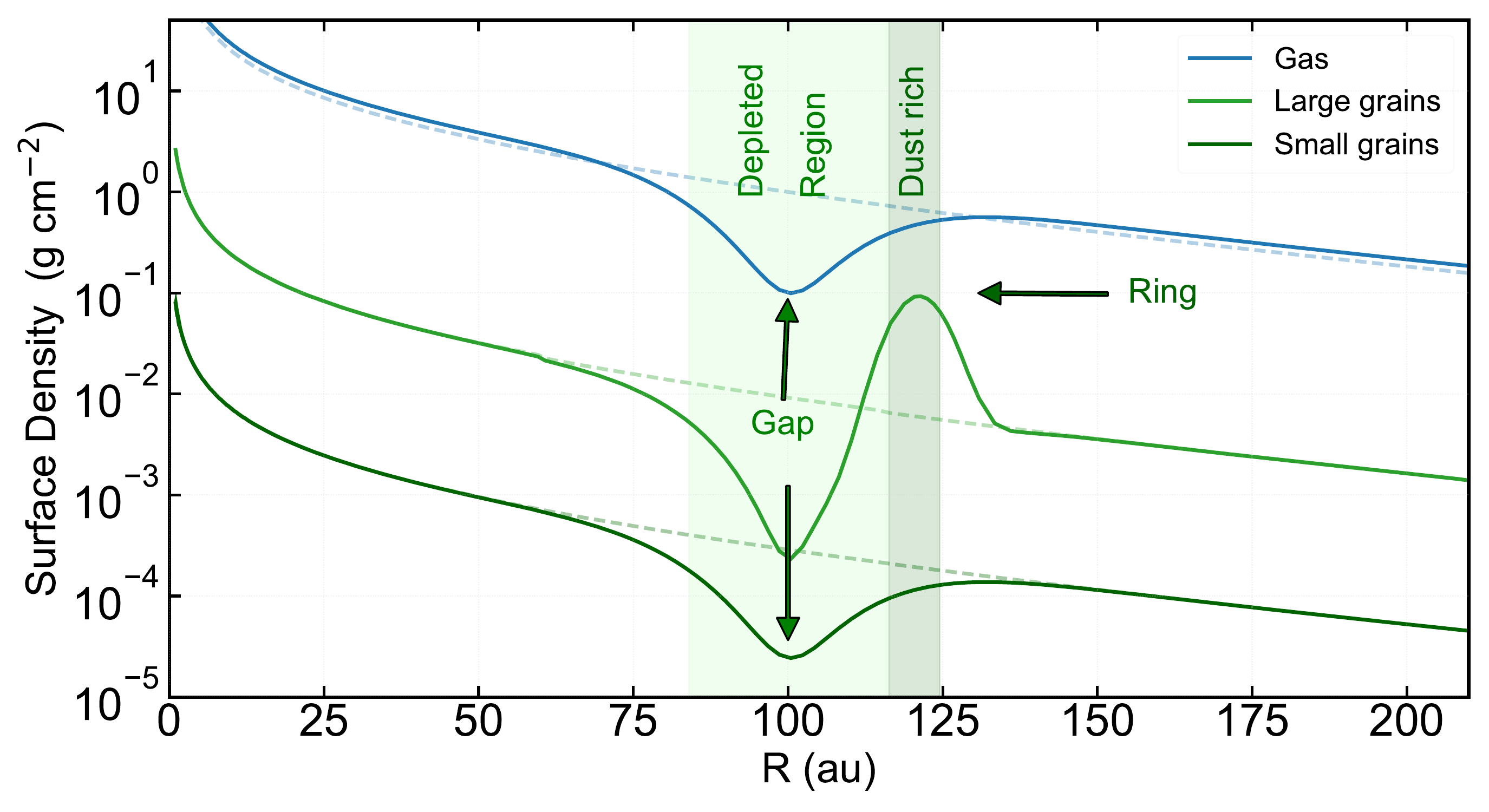}
\caption{Surface density of a modeled disk including the dust substructures. The gap is modeled as a depletion of material in gas and dust, while the ring is an enhancement or dust trap only in the dust large grains. It is noteworthy that the gap has different depletions for large and small grains, as larger grains would be less coupled to the gas. The gap and the ring were modeled using gaussian modulations of the undepleted surface density $\Sigma(r)$, shown with dashed lines. \label{fig:Surf_dens}}
\end{figure}

\begin{figure*}
\includegraphics[width=1.\textwidth]{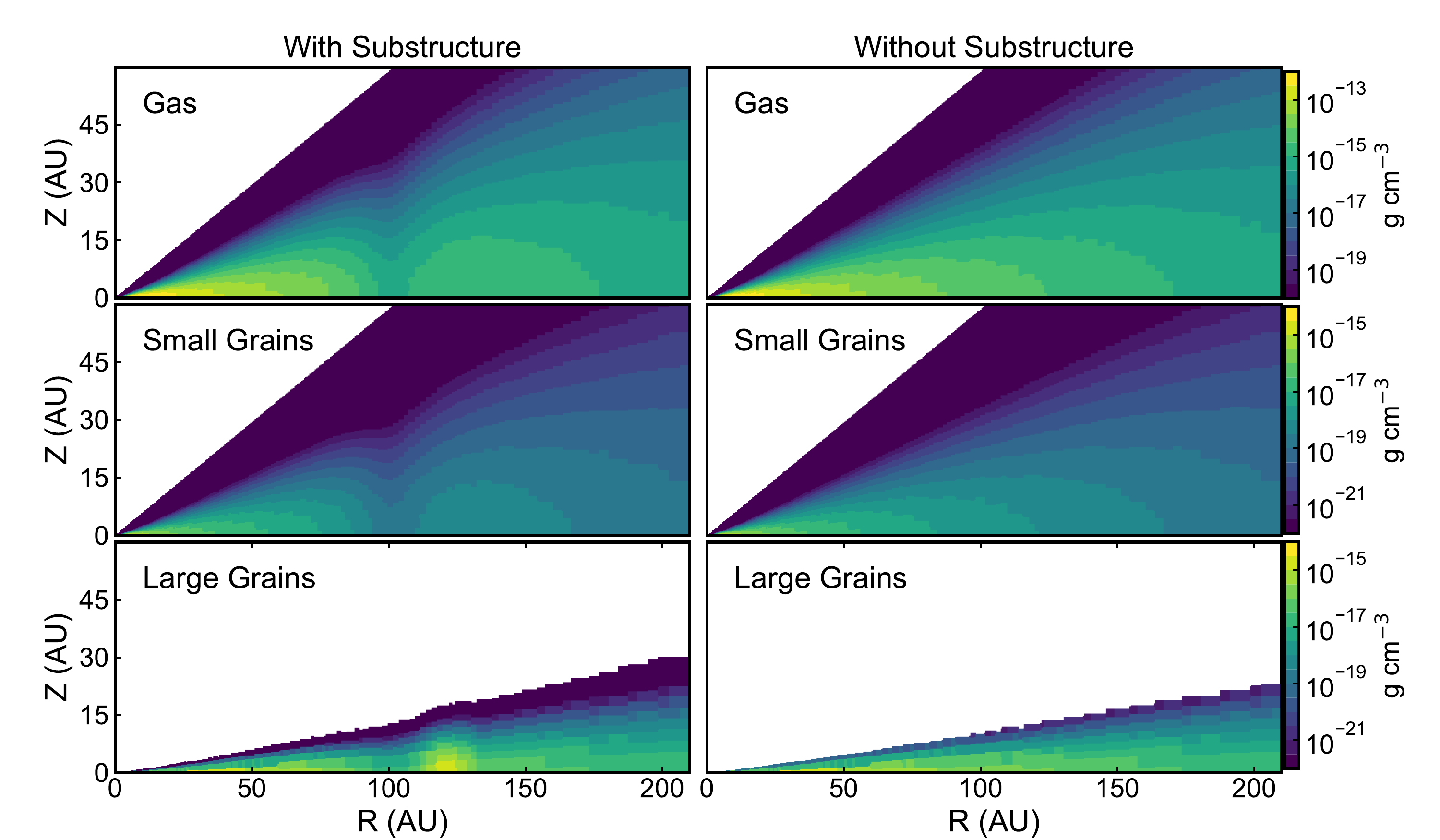}
\caption{2D distribution of gas and dust in the simulations. From top to bottom: Gas, small grains ($a\leq$ 1 $\mu$m) and large grains ($a\leq$ 1 mm) densities. \textbf{Left:} Density field in the disk with the gap-ring substructure. \textbf{Right:} Density fields in the undepleted disk. The gap at 100 au is present in all the three components. However, the ring is only present in the large grains at 120 au. It can also be observed that the large grains are more concentrated in the midplane while the small grains are more coupled to the gas, emulating the dust settling. \label{Fig:Denz_z}}
\end{figure*}

We illustrate how the substructures change the surface density profile  in Fig. \ref{fig:Surf_dens}. We also show the vertical density distribution for the gas and the two main dust populations in Figure \ref{Fig:Denz_z}. The figure shows the dust settling in large grains  and how the gap-ring substructure differs from the basic model. The gap lies at 100 au, and the ring is placed at 120 au.

\subsubsection{Initial Chemical Abundances}

 The initial abundances set the stage for the overall chemical evolution and are given in Table \ref{tab:Abundances}. The elemental abundance for C, H, O, and N is within the range of the reported abundances in the literature for the ISM \citep{Nieva_Przybilla_2012}. At the beginning of our simulations we set CO as the main carbon carrier. The disk shields itself to UV photons with enough energy to photodissociate CO, which implies that at the beginning of the disk, the carbon available will react with the oxygen to create CO \citep{Langer..1976}. We chose to put the CO in the gas phase because the binding energy of CO is low enough for several desorption mechanisms to be effective in keeping the bulk of CO in the gas phase \citep{Bergin..et..al..1995}.
 
 Given that we set the oxygen to be more abundant than carbon in our simulations (C/O $\sim$ 0.4), the remaining oxygen will react to create water. Even though it is pressure dependent, the sublimation temperature of water is above 100 K \citep{Bergin_Cleeves_2018}, which is usually located within the inner few au for the midplane \citep{Lecar..et..al..2006}. So, we choose to input the water as ice, otherwise it would immediately sublimate beyond or above the snowline.
 
 Nitrogen is initially presented in its atomic gas form. The initial form of nitrogen is actually uncertain, but \cite{Schwarz..Bergin..2014} show that nitrogen being initially present in its atomic form is the best match for the current ammonia to water ratio in comets.
 
 Heavy elements are important for the ion balance given that they exchange charge with other molecules and have longer recombination timescales, so they stay charged longer.   Early work suggested that these species are depleted in the dense ISM and we adopt that conclusion as well \citep{Graedel..et..al..1982}. 

\begin{deluxetable}{l|l|l|l}
\tablecaption{Initial Abundances of the simulations with respect to the number of H atoms.\label{tab:Abundances}}
\tablehead{
\colhead{Species} & \colhead{Abundance} &\colhead{Species} & \colhead{Abundance} 
}
\startdata
H$_2$ & 5(-1) & He  & 9(-2)  \\
HD  & 2(-5) &  C  & 1.8(-7)\\
CO & 1.15(-4) & N & 7.5(-5)\\
gH$_2$O & 1.8(-4) & C$^+$ & 1.15(-7) \\
H$_2$O & 1.8(-6) & Si$^+$ & 8(-9)\\
S & 8(-8) & Na$^+$ & 2(-8)  \\
Mg$^+$  & 7(-9) & P & 3(-9) \\
Fe$^+$ & 3(-9) & F & 2(-8)  \\
Cl & 4(-9) & & \\
\enddata
\tablecomments{All the abundances are written in the standard form A(B) = A$\times 10^{B}$}
\end{deluxetable}

\subsection{Image Creation}

 We create synthetic images assuming a distance of 125 pc and a face-on orientation (i=0$^\circ$). The face-on orientation removes systematic degeneration in the ray-tracing and allow us to trace better the emission layer of a given line. We produced spectral cubes containing dust continuum and molecular line emission. We convolved the spectral cubes with circular gaussian beams. We focused on emission falling within Band 6 of ALMA (211 GHz - 275 GHz), observed with an angular resolution of 0."1 (12.5 au at a distance of 125 pc). We generated cubes with a spectral resolution of 0.01  km s$^{-1}$, which are then convolved to a 0.1 km s$^{-1}$ spectral resolution. By using that initial
spectral resolution, we avoid artificial error produced in the image from the ray tracing that samples specific frequencies rather than a frequency bin. We only report the moment 0 maps corresponding to the integrated intensity. Finally, we subtract the dust continuum emission by interpolating a linear function taking values  at the two extremes of the spectral cube.

\section{Results}\label{sec:results}

In our analysis, we compare the difference between the outputs of the fiducial model without any substructure and the model that includes the presence of the gap and the ring. We contrast the physical and chemical structures derived for both models.

\subsection{Physical Conditions}

\subsubsection{Radiation Fields}

\begin{figure*}
\begin{center}
\includegraphics[width=1\textwidth]{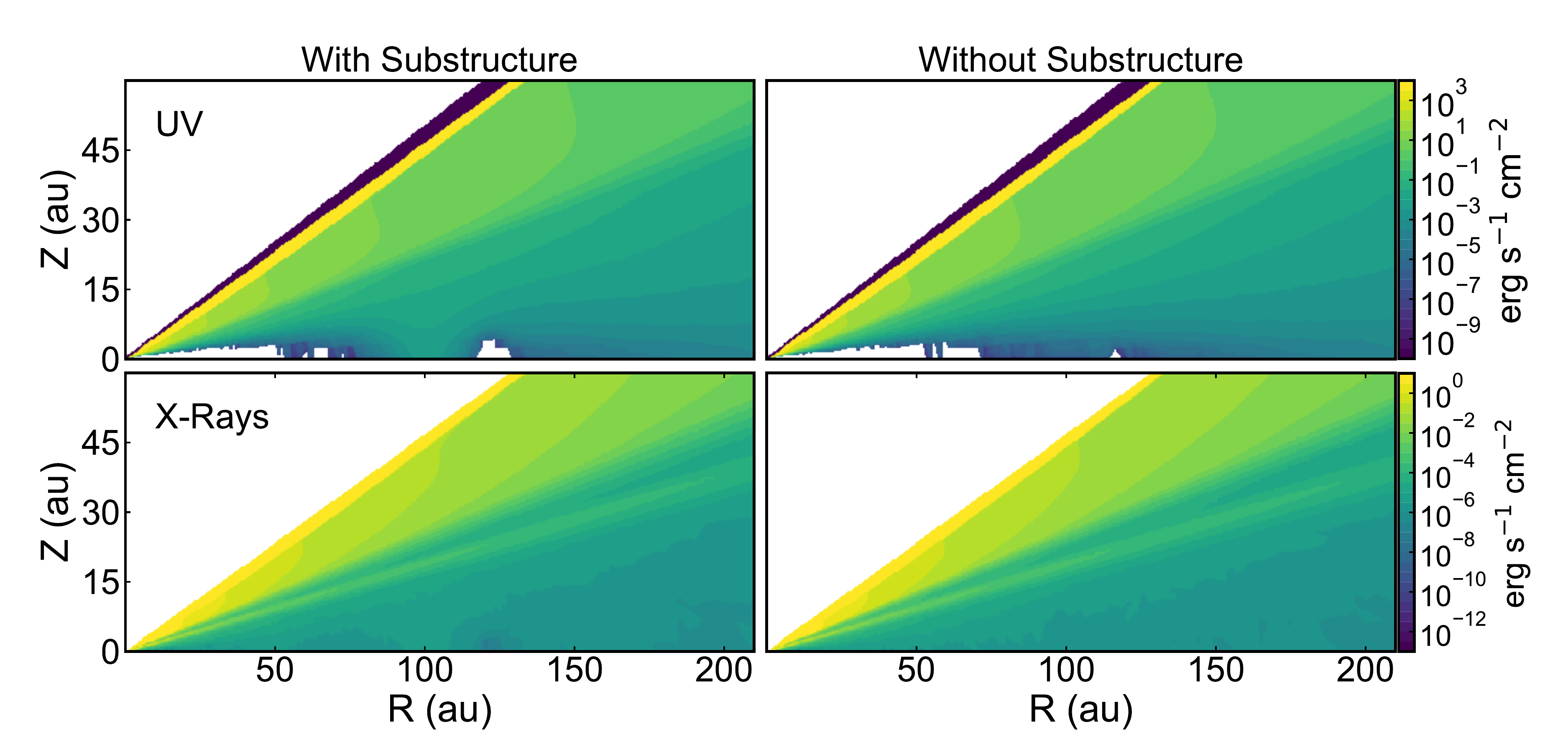}
\end{center}
\caption{Comparison between the UV and X-Rays fields for the model with the dust substructure and the model without the gap-ring substructure. \textbf{Left:} Model including the dust substructure with UV field on top and X-Rays on the bottom. \textbf{Right:} Model without the dust substructure with UV field on top and X-Rays on the bottom. It is noticeable that UV photons penetrate the disk deeper in the gap reaching the layer where planets could be. The UV will not only heat the disk, but they will also modify the chemistry through photodesorption and photodissociation. \label{fig:uv_xray}}
\end{figure*}

\begin{figure}
\begin{center}
\includegraphics[width=.5\textwidth]{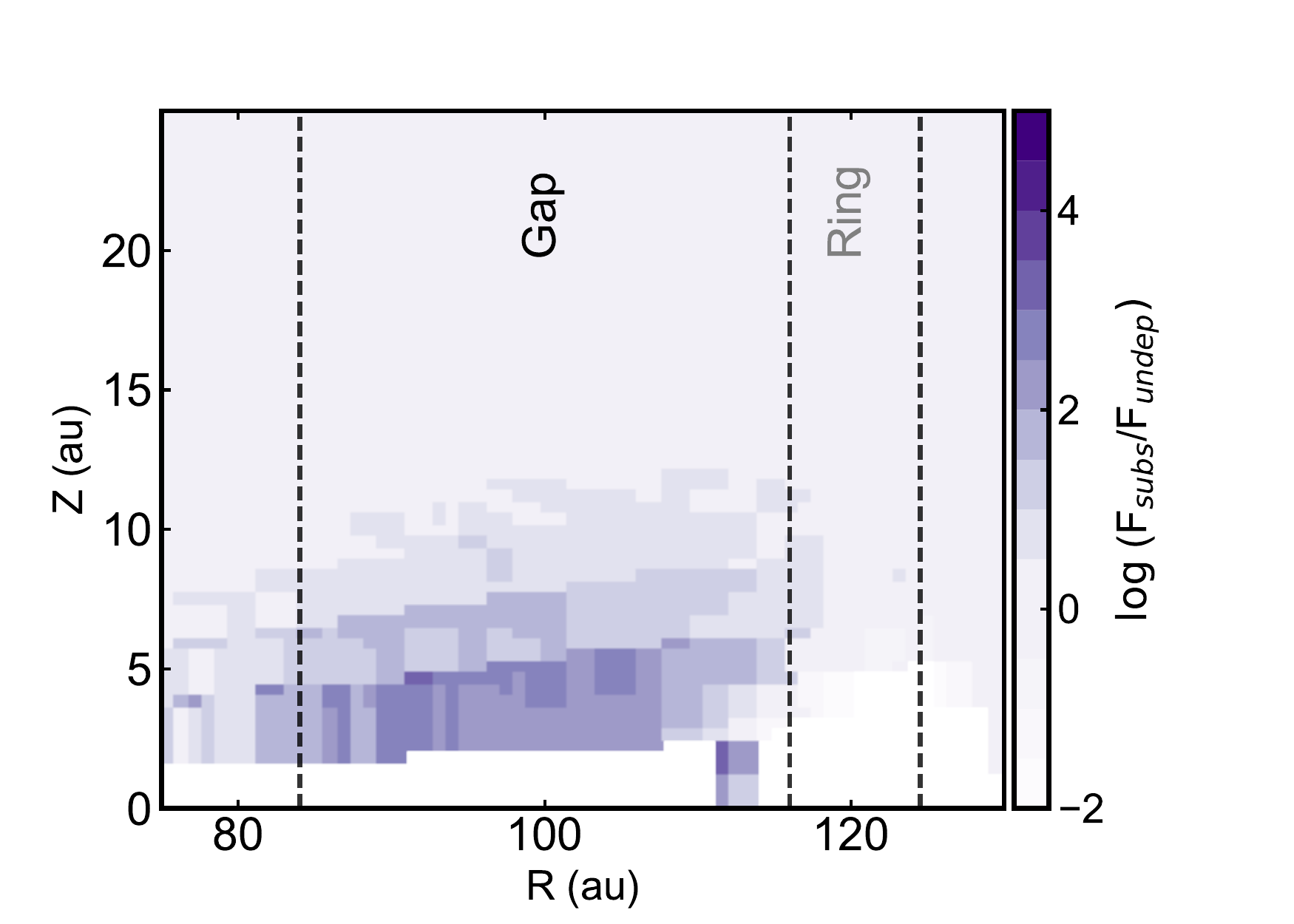}
\end{center}
\caption{Ratio of the UV flux inside the dust substructure that compares the presence of the substructure and the lack of it. The UV ratio is higher and gets deeper in the outer edge of the gap. It is likely caused by back-scattering in the back of the gap that ends up radiating inside the gap. Even though the ratios gets to values of a million times higher, it is noteworthy that the UV flux is already low at that depth (see Fig. \ref{fig:uv_xray} for the flux values).  \label{fig:uv_dif}}
\end{figure}

We show the changes in the radiation fields due to the dust substructure in Figure \ref{fig:uv_xray}. Dust depletion inside the gap decreases the optical depth to high energy photons. UV radiation thus penetrates deeper and closer to the midplane in the gap, and to a less extent X-rays. In contrast, high energy photons cannot penetrate inside the ring due to the high concentration of dust. We zoom-in inside the dust substructure and compare the ratio of the UV fluxes in our two models in Figure \ref{fig:uv_dif}. The figure shows that the UV flux is mostly enhanced in the back of the gap, probably due to backscattering. It is worth mentioning that  the UV flux at deep layers is much lower than on the surface. Thus, even if the UV flux near the midplane is increased by a factor of a million, it does not imply that the UV is actually higher than in the surface. It only states that the UV photons penetrate deeper when a gap has been carved.

The change in X-rays flux is less strong than in the UV. Even though the X-rays flux increases in the gap and in the layer immediately above it, the X-rays ionization rate is low enough that it will not be the main driver of the chemistry in those deep layers (depending on presence/absence of cosmic rays). The shielding effect of the ring in the X-rays is also present as the midplane effectively has zero X-rays flux and ionization. This is because the higher dust density increases the availability of readily absorbing Fe, Mg, Si, and O atoms. This absorption is more closely concentrated towards the midplane than for UV given due to the extra penetration power of X-rays radiation.
\cite{Kim..Turner..2020} show that the X-rays flux increases inside gaps. Our results have a less prominent X-rays flux increase, mainly because their models have a harder X-rays spectrum with X-rays photons at a higher frequency. This has the effect of increasing the ionization level, because the efficiency of the scattering of X-rays by dust grains is proportional to the energy of the X-rays photon \citep{Bethell..Bergin..2011} and there is greater penetration power \citep{Igea_Glassgold_1999}. A harder X-rays spectrum will result in a more significant level of back-scattering from the outer wall of the gap, increasing the X-rays flux within the gap.

\subsubsection{Thermal Structure}

\begin{figure*}
\begin{center}
\includegraphics[width=1.\textwidth]{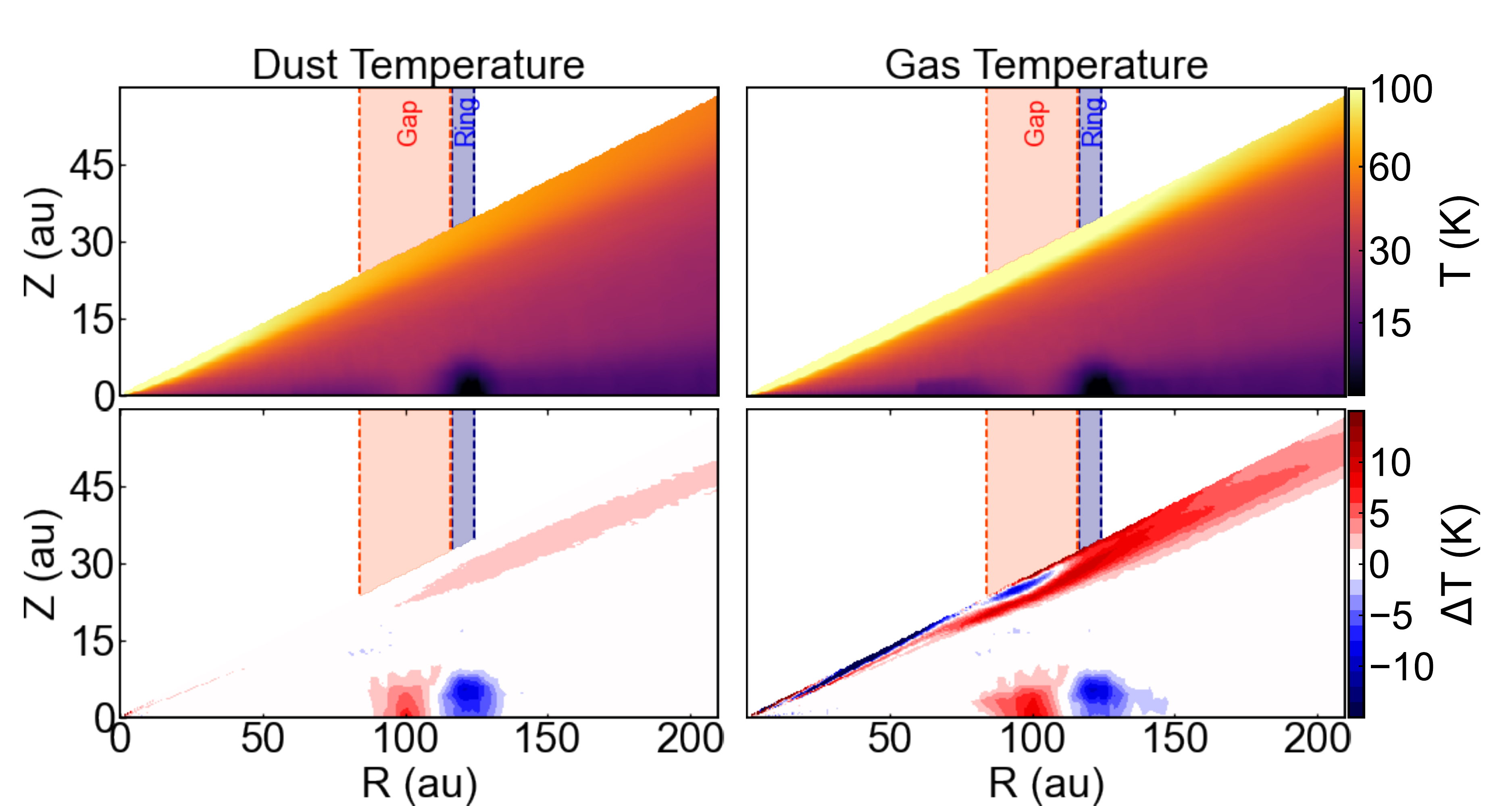}
\end{center} 
\caption{Thermal structure of dust and gas including the gap and the ring in the dust profile. \textbf{Top:} Dust and gas temperature in the disk from left to right. It is observed that the ring is noticeable colder, while the gas is particularly hotter in the gap. \textbf{Bottom:} Difference in temperature between the disk model with the dust substructure and the same disk model but without the substructure. It shows that the gap in both cases could be hotter up to more than 5 K, while the ring is much colder, by almost 10 K.}
\label{Fig:Temp_2D}
\end{figure*}

The thermal structure of the gas and the dust have significant changes in the gap and ring. \cite{Teague..2017}, \cite{Vandermarel..et..al..2018} and \cite{Facchini..et..al..2018} have looked at the thermal fields inside the dust substructure, although getting different results; such differences will be discussed further in Section \ref{sect: Comparison}. In our models, there is a noticeable cooling effect  at the ring location, where the gas and dust temperature decrease by more than 5 K. That slight change becomes significant when compared with the expected temperature from the basic model, which is slightly below 20 K. On the other hand, the dust temperature is slightly increased in the gap. In Figure \ref{Fig:Temp_2D}, we take the difference between our two models to isolate the change induced within the gap/ring.  Here we observe a dust temperature increase of almost 5 K within the gap and a decrease of 10 K in the ring.
 The effect is caused  by the change in the overall opacity to the stellar irradiation and high energy photons. Inside the gap, the optical and UV photons will reach deeper and closer to the midplane, having a net heating outcome. The opposite effect is observed in the ring, where dust grains shield themselves and the gas to the general radiation, particularly in the midplane, meaning that the optical depth is increased, so optical and UV-photons are not able to reach inside the ring.
 
 In the upper layers of the atmosphere, we also observe a slight increase in temperature. We associate it with the fact that photons in the upper atmosphere are able to reach farther distances in the disk, due to the depletion of material in the gap, which reduces the line of sight optical depth, measured form the star.

\begin{figure}
\includegraphics[width=0.5\textwidth]{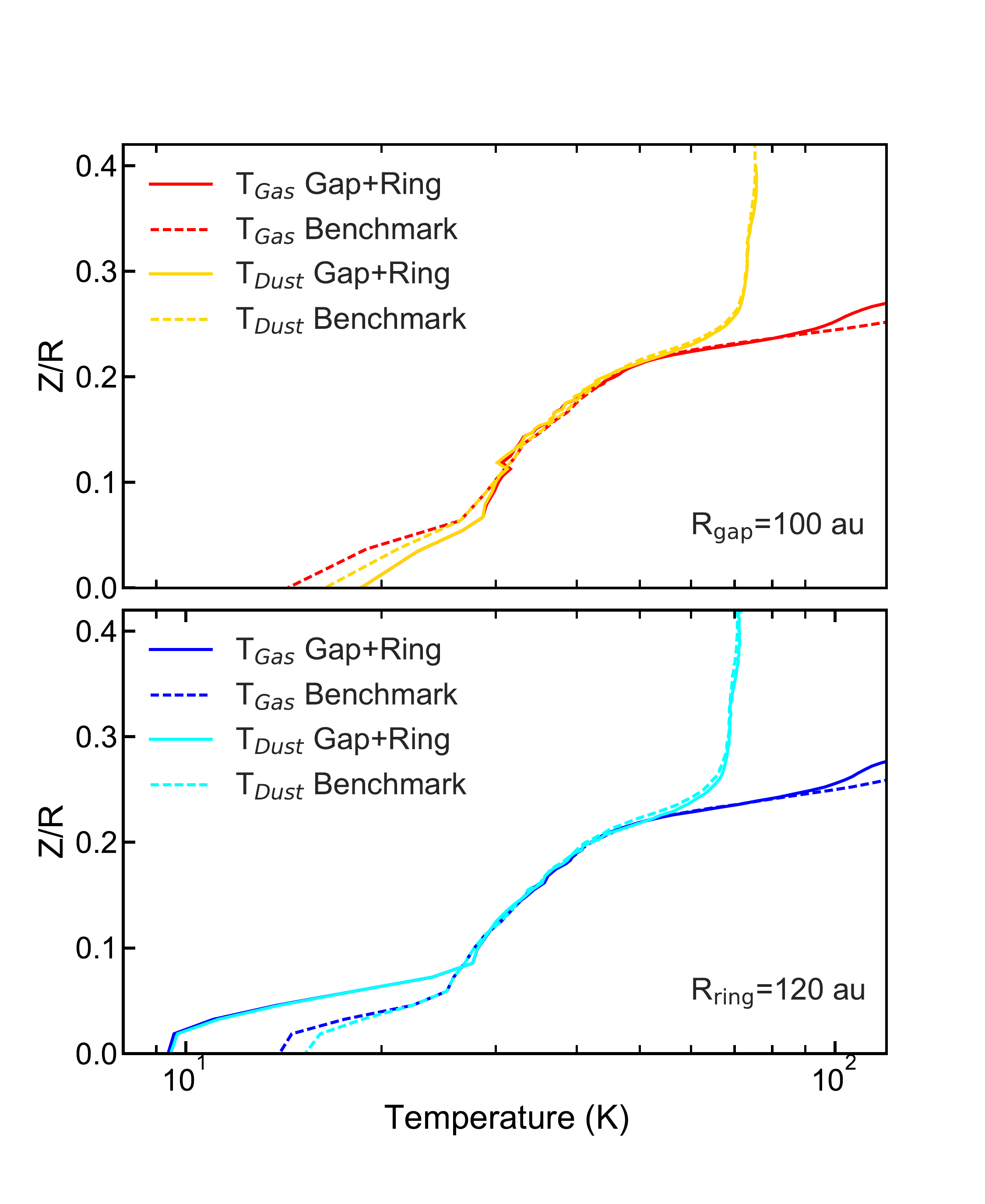}
\caption{The figures illustrate a vertical cut of the thermal field in the middle of the gap and the ring comparing it with a model without them. \textbf{Top:} Vertical cut at 100 au in the center of the gap.
\textbf{Bottom:} Vertical cut at 120 au in the center of the ring. They both show that the change in temperature happens within one scale height ($z/r< 0.08$), where the rings gets colder by more than 5 K. Even if the increment in temperature in the gap is smaller it could be enough to thermally desorbed molecules like CO.}
\label{Fig:Temp_cuts}
\end{figure}

The disk is optically thin to UV and optical radiation in its higher layers. However, the significant changes to the optical depth occur deeper in the disk, closer to the midplane as it is observed in Figure \ref{Fig:Temp_cuts}. The Figure shows vertical thermal cuts in the disk for both, the gap and the ring and how they diverge at lower heights. The thermal influence of the gap-ring substructure in the dust is produced closer to the midplane ($z/r<0.1$), where the dust mass is concentrated, rather than the dust-depleted disk surface.  In terms of the net temperature variation, the thermal change in the ring is more significant than the increase of temperature in the gap.

\subsection{Chemistry in Gaps and Rings}

\subsubsection{Radial Abundance Structure}

\begin{figure*}
\begin{center}
\includegraphics[width=1\textwidth]{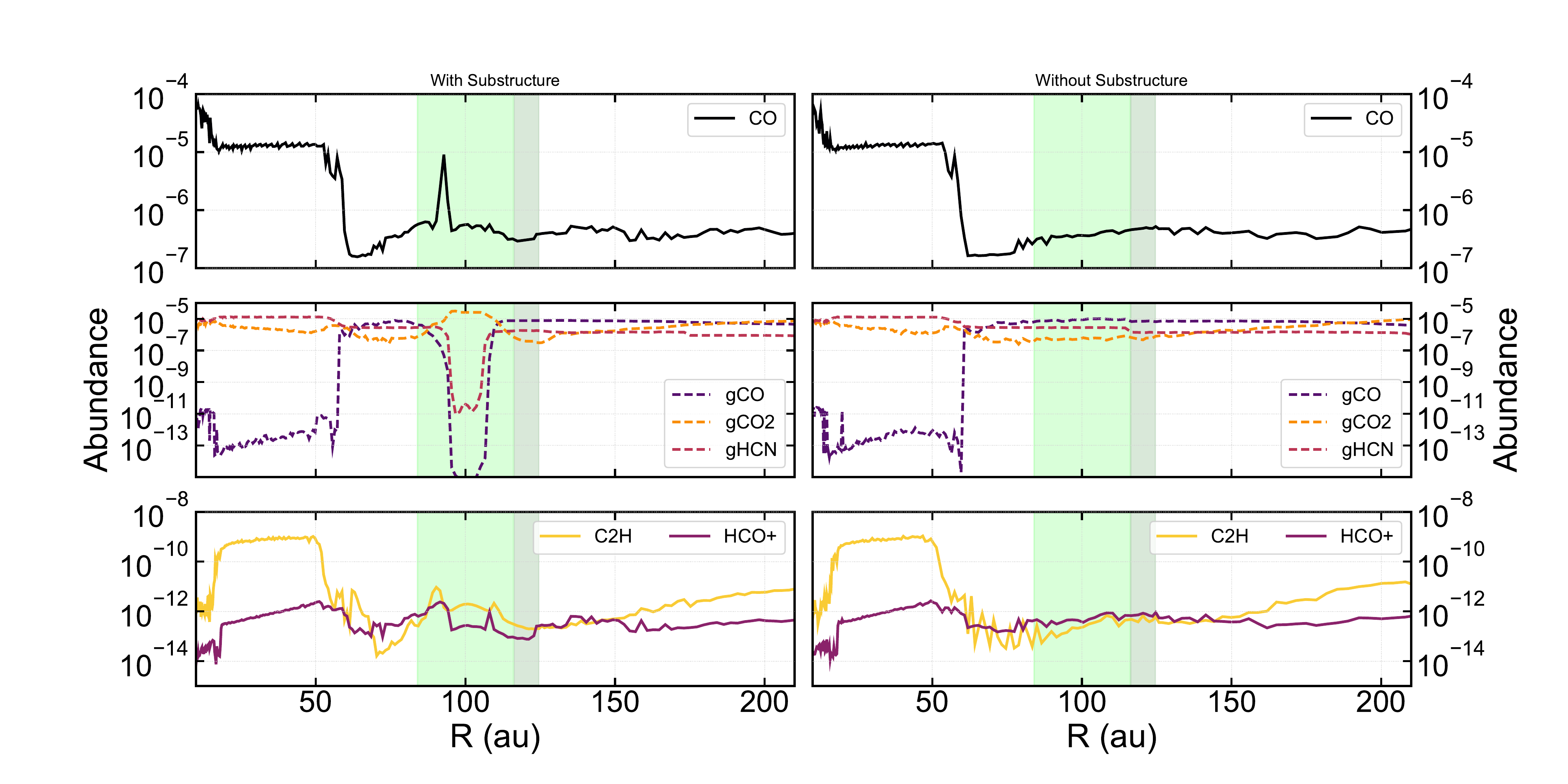}
\end{center} 
\caption{Abundances radial profiles for different species of interest in a disk with the gap and the ring (left) and the benchmark without them (right). \textbf{Top:} CO radial profile showing its sublimation into the gas phase inside the gap. CO abundance increases by an order of magnitude at the inner edge of the gap because of CO thermal desorption. \textbf{Middle:} \textbf{Carbon carriers in dust grains in the ring and the gap. It is noteworthy the shifting between CO and CO$_2$ in the grains caused by the water dissociation and the subsequent reaction between CO and the OH radical.} \textbf{Bottom:} Photochemical tracers that show the change in opacity to high energy photons in the gap and in the ring. }
\label{Fig:Abundances Profiles}
\end{figure*}

\begin{figure*}
\begin{center}
\includegraphics[width=1.\linewidth]{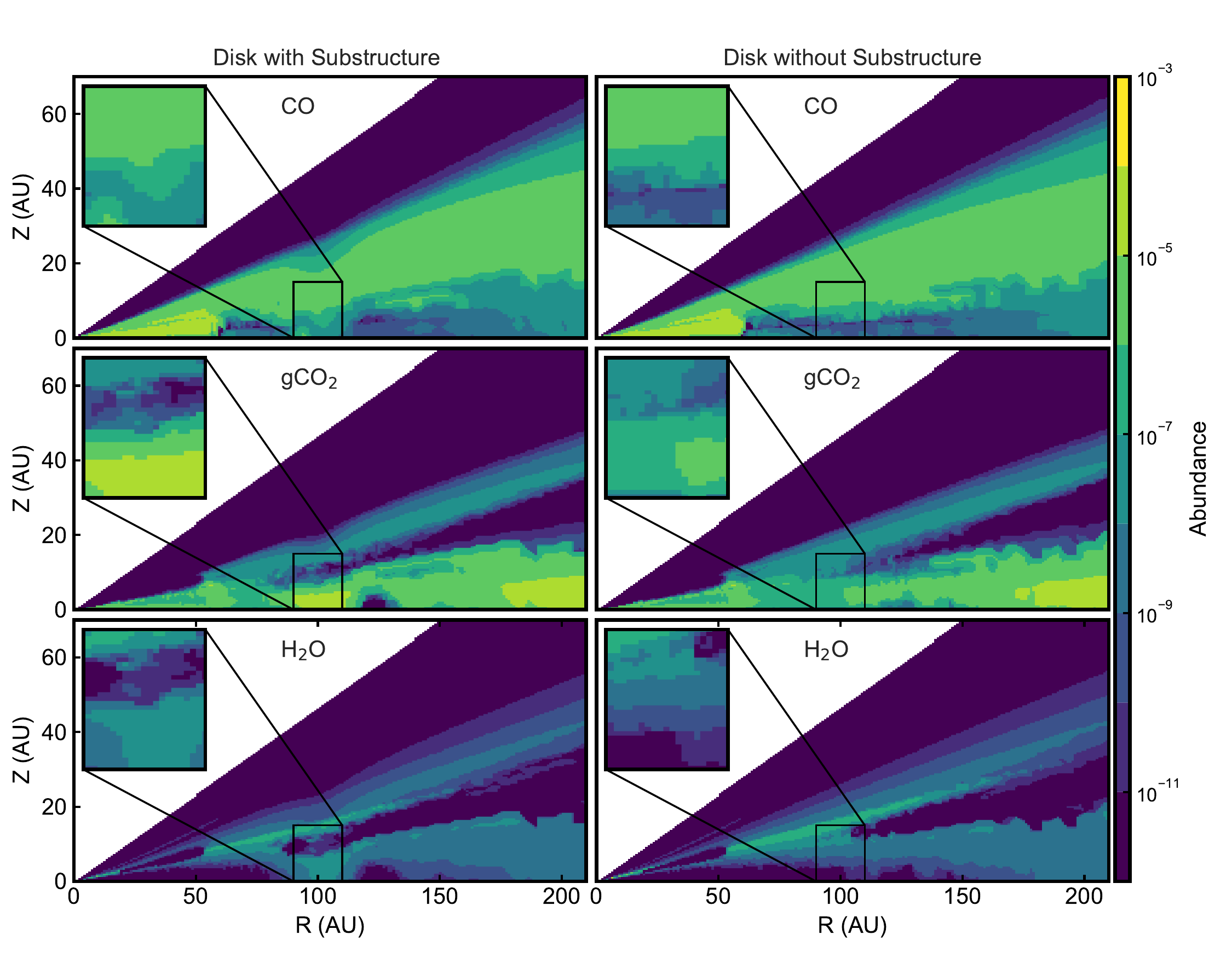}
\end{center} 
\caption{Comparison of the abundances of certain species of interest for a disk with the gap-ring substructure and the benchmark. \textbf{Left:} 2D abundances in the disk model including the dust substructure. \textbf{Right:} 2D abundances in the disk without dust substructure.  The zoom-in shows the gap location. The zoom-in illustrates the CO sublimation in the gap, \textbf{in particular at the inner edge of the gap}. It also shows the production of frozen CO$_2$ in the grains right above the midplane for the model with the dust substructure. Water is being desorbed and photodissociated, which liberates an OH radical that will oxygenate the chemistry. Overall, the gap and the ring have the power to increase or decrease the abundance or molecules of interest in orders of magnitude, which should be taken into account for the observation of line emission.}
\label{Fig:Abundances2D}
\end{figure*}

We show the different profiles comparing the impact of dust substructure in Figure \ref{Fig:Abundances Profiles}. Each abundance was obtained by dividing the column density of a given species by the total number of gas particles at a given radius, i.e., 

\begin{equation}
    X_i(r) = \frac{\int X_i(r,z) n(r,z)dz}{\int n(r,z)dz},
\end{equation}

\noindent with $X_i(r,z)$ the abundance of a given species at a given $r$ and $z$, and $n(r,z)$ the gas number density field.

If we focus only in CO, we observe that the CO ice sublimates in the gap, but there is also chemical reprocessing, such that the carbon in the gas phase within CO is transferred into CO$_2$ ice on the grains. Thus, the C atoms that would be frozen on the grains are now divided between volatiles species (CO) and the grains (CO$_2$), even for a gap located at 100 au. Inside the gap, the volatile CO increases its abundance by more than one order of magnitude, fully compensating for gas depletion inside the gap in this case. However, there is an interplay between the depth of the gap and its physical conditions. This leads the CO abundance increase to be concentrated near the midplane and not distributed vertically within the gap, as seen in Fig. \ref{Fig:Abundances2D}. In this particular model the gap is located beyond the CO sublimation front, so CO will be predominantly found in the warm molecular layer of the gap that is above the vertical CO snowline \citep{Aikawa..et..al..2002}. CO sublimation in the gap occurs close to the midplane driven by thermal and photodesorption, making the CO to be more concentrated below the vertical CO snow line of the undepleted case, as it is shown in Fig. \ref{Fig:Abundances2D}.

The CO in the ring does not show strong trends, although the carbon gets transferred to different carriers. The frozen CO does not show strong changes; however, hydrogen cyanide (HCN), a nitrogen compound, becomes a main carbon carrier. The main carbon carriers and the related chemistry  are presented in Sect. \ref{Sect: CC}.  The middle panel in Figure \ref{Fig:Abundances Profiles} illustrates how the dust substructure shifts the carbon carriers in the dust grains. The main carbon carrier in the midplane at 100 au shifts from CO to CO$_2$, where the CO in the grains gets oxidized to create CO$_2$, to the detriment of the CO abundance. Nevertheless, in the very center of the gap, where the temperature is the highest and the optical depth the lowest, gas phase CO still remains the dominant carbon carrier. The increased rate of photodissociation, higher temperature and lower density counteract the formation of CO$_2$. 

The last panel in Fig. \ref{Fig:Abundances Profiles} shows photochemical tracers in the disk, C$_2$H and  HCO$^{+}$. C$_2$H is an important tracer of the UV-driven chemistry and also of the $\tau_{\rm UV}=1$ layer \citep{Henning..et..al..2010}. C$_2$H also responds to the CO distribution, such that it is abundant right above the CO self-shielding layer. 
We would expect that since the penetration of UV photons is deeper in the gap, the abundance of C$_2$H, a photochemical tracer, might be enhanced. Figure \ref{Fig:Abundances Profiles}  shows a peak in the C$_2$H and HCO$^{+}$ abundances inside the gap. The increase in C$_2$H abundance associated with the gap is compensated by the depleted gas surface density. Such interplay between abundance and surface density causes C$_2$H  to have a broad flat distribution - notably not a decline in emission given the drop in the overall surface density.  The predicted vertically averaged abundances in this model are quite low, and below what is inferred from observations of $\sim$10$^{-7}$ \citep{Bergin..et..al..2016, Bergner..et..al..2019}. The UV field inside the gap is favourable for the production of C$_2$H. However, in order to increase the C$_2$H abundance, a C/O ratio higher than unity is needed, otherwise CO will be carrying most of the carbon content in the substructure \citep{Bergin..et..al..2016,Kama..et..al..2016,Miotello..et..al..2019}. Therefore, a carbon source inside the gap is necessary to locally increased the C/O ratio, which in turn will enhance the C$_2$H abundance. Some possible carbon sources in the substructure will be further discussed in Sections \ref{ssection: MF} and \ref{ssection: Limits}. 

In addition to C$_2$H, we show HCO$^{+}$, which forms via X-rays  ionization on disk surfaces \citep{Semenov..2004,Cleeves..et.al..2015}. Regarding a photochemical tracer more sensitive to X-rays, HCO$^{+}$ abundance also increases in the center of the gap. The HCO$^{+}$ increase is due to the CO sublimation right in the center of the gap, which is one of the main species that could lead to the formation of HCO$^{+}$ through the reaction of H$_2^{+}$ and CO. In the colder outer regions, even though X-rays and cosmic-rays have a higher mean free path, HCO$^+$ is less abundant because it recombines transferring the carbon to other molecules such as CO, CO$_2$ or H$_2$CO.  

In general, we find that the ring at 120 au is not favorable for the production of photochemical tracers. The dust enhancement within the ring shields the midplane from high energy radiation and ionization. However, the ionization rate in the midplane is already low in the undepleted case, the addition of a substructure produces an even lower abundance of photochemical tracers. Therefore, the ring does not have an observable effect in the disk photochemistry. However, the ring is colder and depending on the location with respect to the star may provide an enhanced site for molecular freeze-out \citep[e.g.,][]{Krijt_et_al_2016} which leads to localized ice enhancements.

\subsubsection{Vertical Shift of the Warm Molecular Layer}

\begin{figure}
\includegraphics[width=0.5\textwidth]{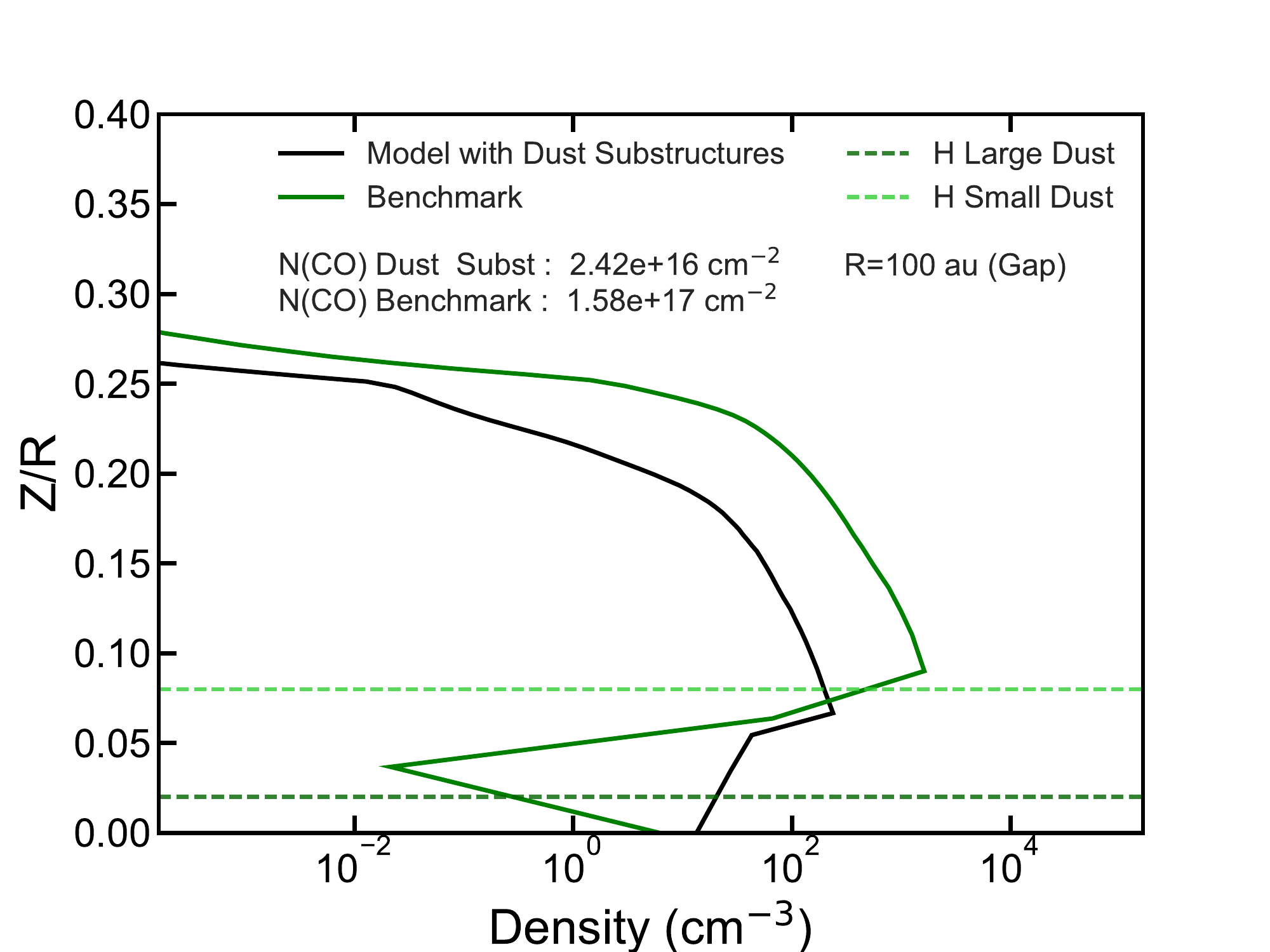}
\includegraphics[width=0.5\textwidth]{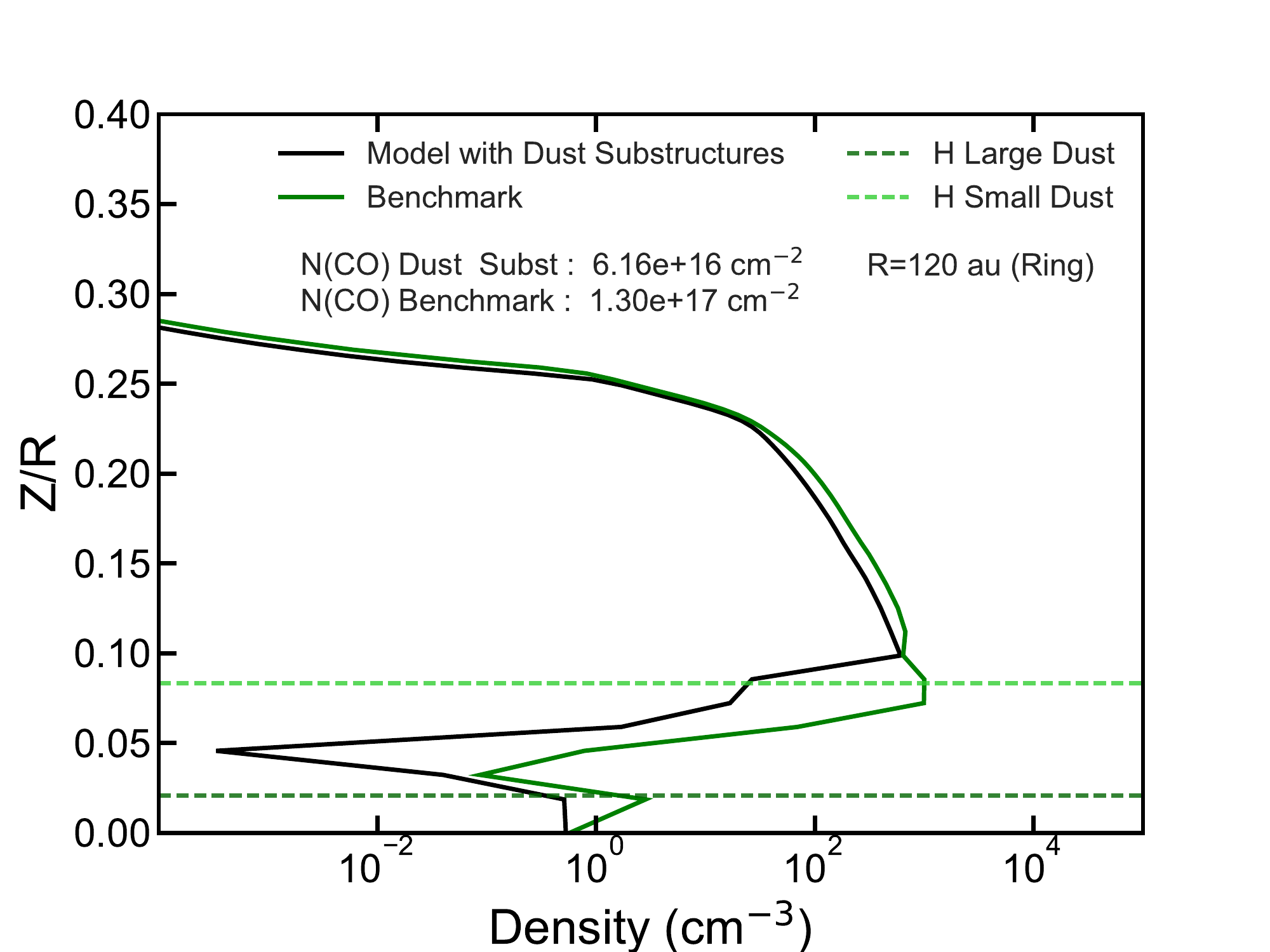}
\caption{We show a vertical cut of the CO density in the middle of the gap and the ring in comparison with the benchmark model. \textbf{Top:} Vertical cut at 100 au in the center of the gap.
\textbf{Bottom:} Vertical cut at 120 au in the center of the ring. Inside the gap it is important to consider that the disk has been depleted by a factor of 92.5$\%$ and 97.5$\%$ for small and large grains respectively. Due to a higher photodesorption rate, the CO is mostly gaseous in the midplane rather than frozen out. The ring on the other hand is not material depleted and the CO will get frozen-out right between the scale height if the small and large grains. We observe that the CO column densities do not scale with the local surface density once the thermo-chemical effect of the substructure has been considered}. Overall, the chemical substructures will have a more significant influence in the midplane chemistry rather than at the surface.
\label{Fig:WML}
\end{figure}

We observe that with the inclusion of a gap in the gas and dust the disk becomes `chemically thinner'. It is chemically thinner in the sense that the warm molecular layer has a vertical shift to a lower layer closer to the midplane, as it can be observed in Fig. \ref{Fig:WML}. The warm molecular layer \citep{Aikawa..et..al..2002} is the visible gas phase chemistry layer in the disk where CO sublimates at T$_{\rm dust}\ >$ 20 K \citep{Bergin_Cleeves_2018}, and the UV optical depth exceeds unity, i.e.,  molecular photodissociation becomes less important.  This layer generally resides in the disk surface layers as seen in our fiducial model (Figures \ref{Fig:Abundances2D} and \ref{Fig:WML}). In the case of a disk with a gap, this layer shifts deeper in the disk closer to the midplane as shown in Fig. \ref{Fig:WML}. This suggests that there is a chemically active layer closer to the expected location of protoplanets driving the gap formation. Under strong levels of  vertical mixing, the shift could be significant enough to ease the transport of volatiles between the warm molecular layer and the protoplanet at the disk midplane.

At the location of the ring, given that the disk is colder, most of the molecules are already frozen. Therefore, there is no apparent shift of the warm molecular layer. However, if the ring were located closer to the star, it is possible that this conclusion could change. The role of the ring for terrestrial seeds, would be the opposite to the giant planet that carved the gap. The warm molecular layer could potentially be pushed to higher altitudes.  The colder temperatures in the ring would then enhance the freeze-out of certain species in a location that is potentially the next site of planetesimal formation.

\subsubsection{Carbon Carriers}\label{Sect: CC}

\begin{figure}
\includegraphics[width=0.5\textwidth]{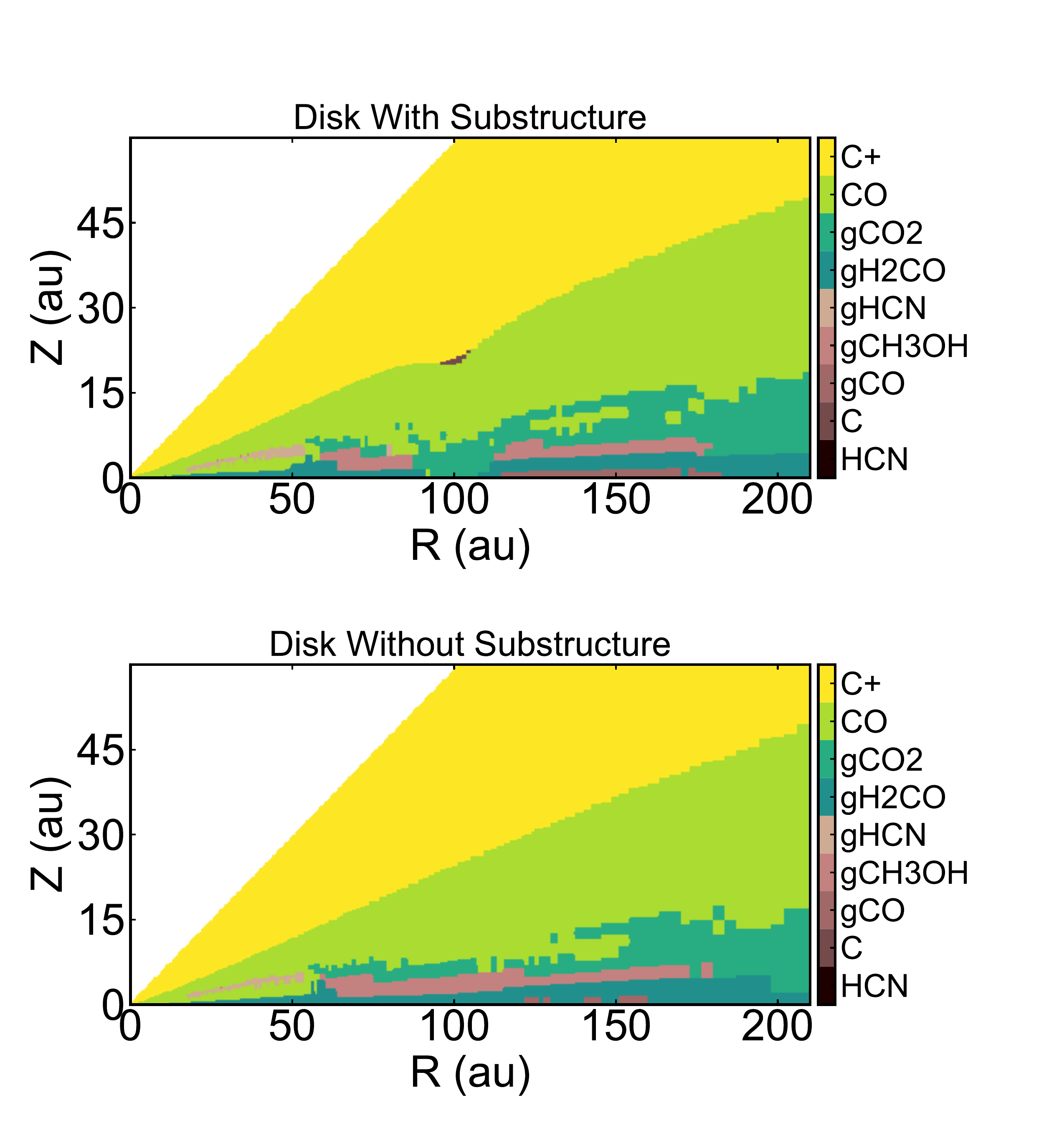}
\caption{Comparison of the species carrying the highest amount of carbon atoms at different locations of the disk. \textbf{Top:} Disk model including the gap-ring substructure. \textbf{Bottom:} Disk model without the substructure. The C$^+$/C/CO layered structure of the PDR illustrates the vertical changes that the dust substructure causes. The ring and the gap will shift up or down the layers were a given carbon carrier will be dominant. We observe that CO$_2$ and HCN become dominant C-carriers in the gap and the ring respectively due to thermo-chemical effect, while the CO layer gets pushed down towards the midplane inside the gap.}
\label{Fig:C_carriers}
\end{figure}

We show the main carbon carriers (species carrying the highest number of carbon atoms) at each location in Fig. \ref{Fig:C_carriers}. The primary change is the enhanced presence of CO closer to the midplane within the gap (i.e the shift of the warm molecular layer).   This also produces a localized change in the C+/C/CO transition zone. 

Regarding the solid ices within the disk, an important phenomenon is the photodissociation and photodesorption of water molecules through UV field induced by cosmic-rays, which takes place in the midplane. The photodissociation of water inside the gap generates an OH radical that reacts with CO on the grains forming CO$_2$ ices. Therefore, CO$_2$ becomes the main carbon carrier at those layers. However, this outcome is dependent on the local C/O ratio and water being dissociated to oxygenate the medium. The enhancement of CO$_2$ is seen in Figure \ref{Fig:Abundances Profiles} and Figure \ref{Fig:Abundances2D}.

 Inside the ring, instead of H$_2$CO being the main carbon carrier at that region, as seen in the baseline model, we see a significant amount of C and N within HCN. To make HCN the main carbon/nitrogen carrier, the chemical path taken is trough the reaction between  frozen nitrogen and CH. The nitrogen comes from a gas phase reaction of the ammonia cation, NH$^+$, with molecular hydrogen that freezes-out; while the CH is produced by photodissociation of small organics. In other words, the low temperature in the ring, along with cosmic-rays, allows the sublimation of nitrogen and the  photodissociation of small organic molecules, producing HCN on the surface of the dust grains.

\subsection{Observational Predictions}
The sublimation of CO in the midplane inside the gap will change the emission of CO and its isotopologues. If the gap is not deep enough, CO emission changes from an optically thin regime in an undepleted disk to an optically thick regime in the disk with the gap, an effect that has been reported by \cite{Vandermarel..et..al..2019}  and \cite{Facchini..et..al..2018}. However, whether or not the CO emission is increased locally is be dependant on the excitation conditions in the gap, specifically the local density and temperature. 

\begin{figure*}
\begin{center}
\includegraphics[height=.3\textheight]{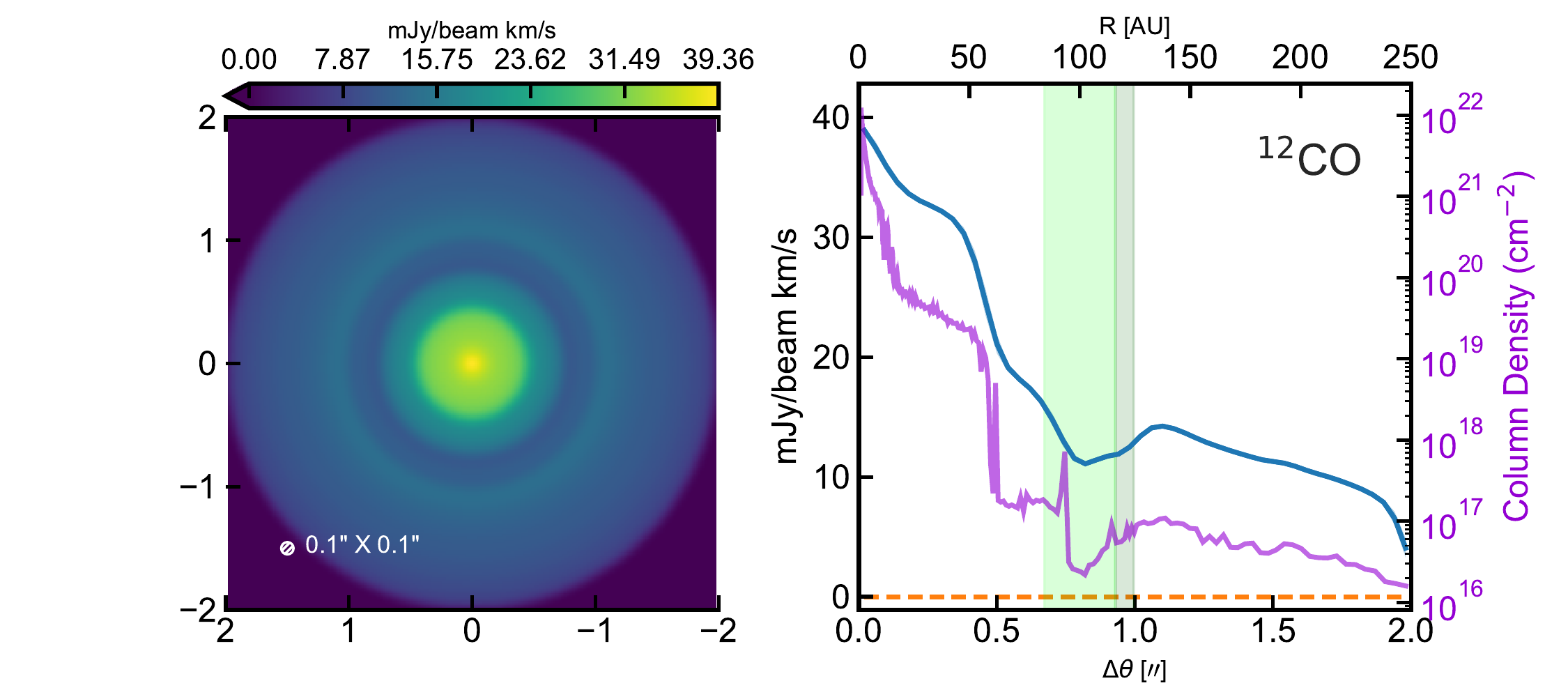}
\includegraphics[height=.3\textheight]{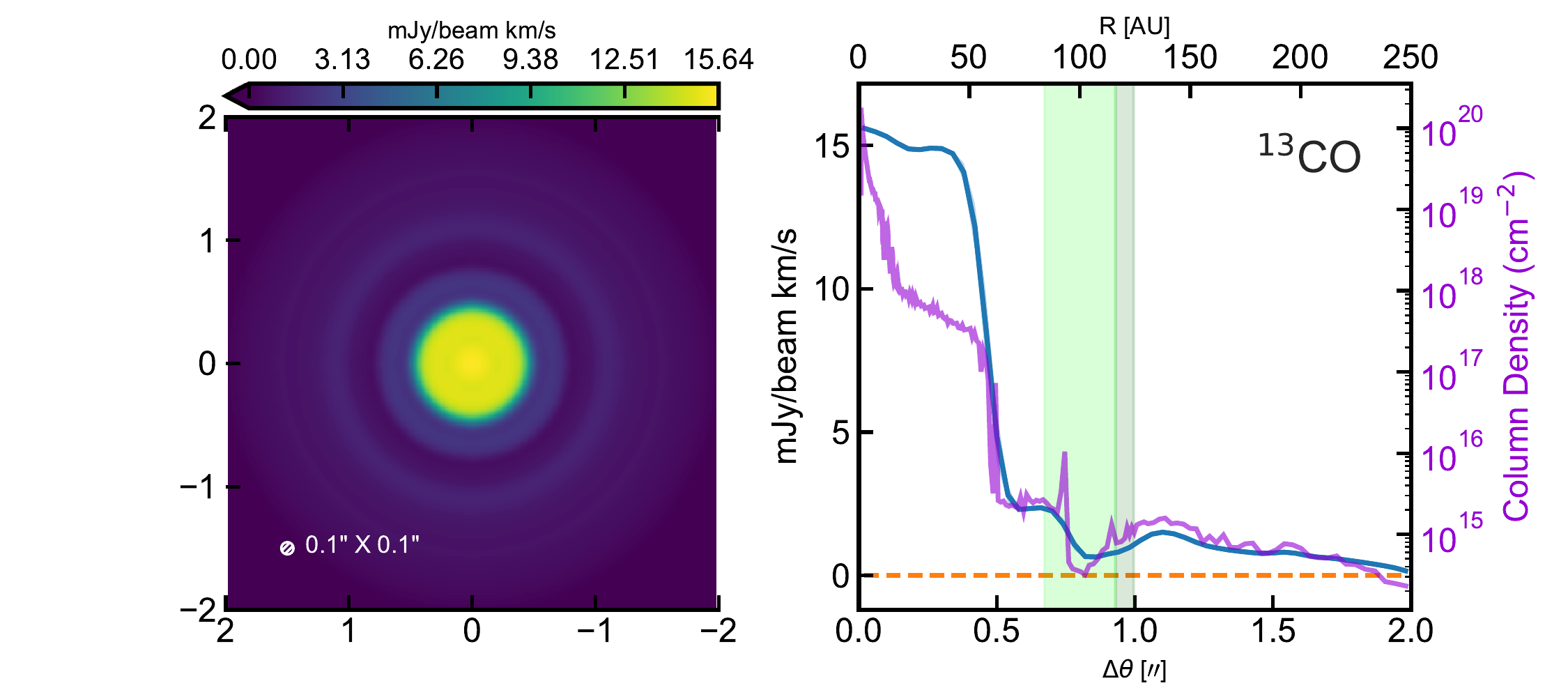}
\includegraphics[height=.3\textheight]{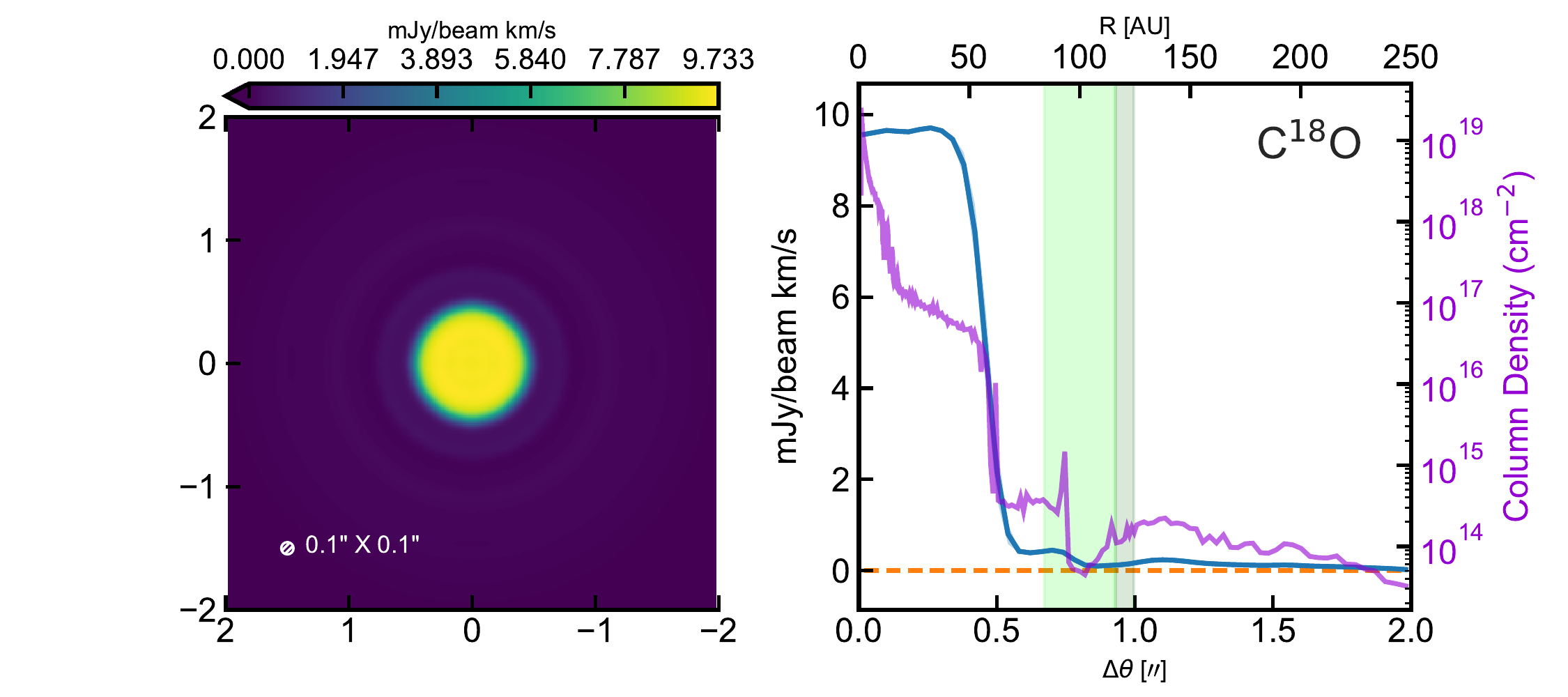}
\end{center}
\caption{Synthetic images for CO J=2-1 transition with 0.1 arcsec resolution. We show the radial profile of the integrated intensity in blue and the radial profile of column density for the respective CO isotopologue in purple. \textbf{Top}: $^{12}$CO. \textbf{Middle}: $^{13}$CO. \textbf{Bottom}: C$^{18}$O. The radial profiles show a dip in the gap for $^{12}$CO, the optically thick tracer. On the contrary, the emission peaks for the optically thin ones, and then it decrease in the ring. In particular for CO in the gap, it will be concentrated in the midplane, which is colder than the upper layers so its emission will be lower for $^{12}$CO, which is optically thick, but it will be higher for $^{13}$CO and C$^{18}$O due to the fact that they are in the optically thick regime. However, because the increase in CO abundance is coming from deeper and colder layers of the disk, the increase CO emission is small. The different consequences of the substructures in the line emission of CO shows a degeneracy between the column density (disk mass+abundance) and the depth of the substructure.}
\label{Fig:syn_images}
\end{figure*}

 We create synthetic images for the CO and its $^{13}$CO and C$^{18}$O isotopologues in the J=2-1 transitions at 0.''1 resolution, observable in Band 6 of ALMA (211 GHz - 275 GHz). We show the integrated intensity images and their respective CO column density and emission radial profiles in Figure \ref{Fig:syn_images}. Our predictions show that for a gap with the chosen depth, the emission at the location of the dust substructure is lower, but the overall emission does not correlate with the depth of the gap or the CO column density. The gas depletion is 92\% of the undepleted value, i.e., $0.08\cdot\Sigma_{\rm basic}(r)$; but even if the CO emission does decrease significantly, it is less than a factor of 2.   In this regard, there are two effects that are relevant towards the emission as the higher temperature tend to increase emission in specific rotational states, while the gas density decrease in the gap would lower emission.   In order to disentangle the effects of gas density  and temperature on CO emission, the thermal structure of the disk is needed. This effect has been previously discussed by \citet{Facchini..et..al..2018} in a gap in the inner disk at $\sim$20 au.   In our model, for  optically thin lines, the depth of the gap in emission is less than that within the H$_2$ gas itself. CO sublimation inside the gap counters gas depletion, which is particularly relevant beyond the CO snowline.

Among the different CO isotopologues, the less abundant isotopologues are better tracers of the gap chemistry. Given that they have relatively lower optical depth, they trace deeper layers closer to where the CO desorption front is located in the gap. However, the gas temperature in the midplane of the gap, even if elevated, remains colder than in the upper layers surrounding the gap.  To distinguish these differences requires high resolution and sensitive observations.

\subsubsection{Effect of Disk Masses}

In order to test the possible effect of disk mass  on the overall chemical distributions, we explore a simulation that increases the gas and dust mass by a factor of ten.  In this model we keep the same relative gap depth and ring enrichment. Even though the actual material depletion is different in absolute terms, the gaps carved by planets are dependent on the host disk properties and will scale with the surface density profiles \citep{Duffel_MacFadyen_2013,Fung_et_al_2014,Duffell_2015}.

\begin{figure}
\includegraphics[width=.45\textwidth]{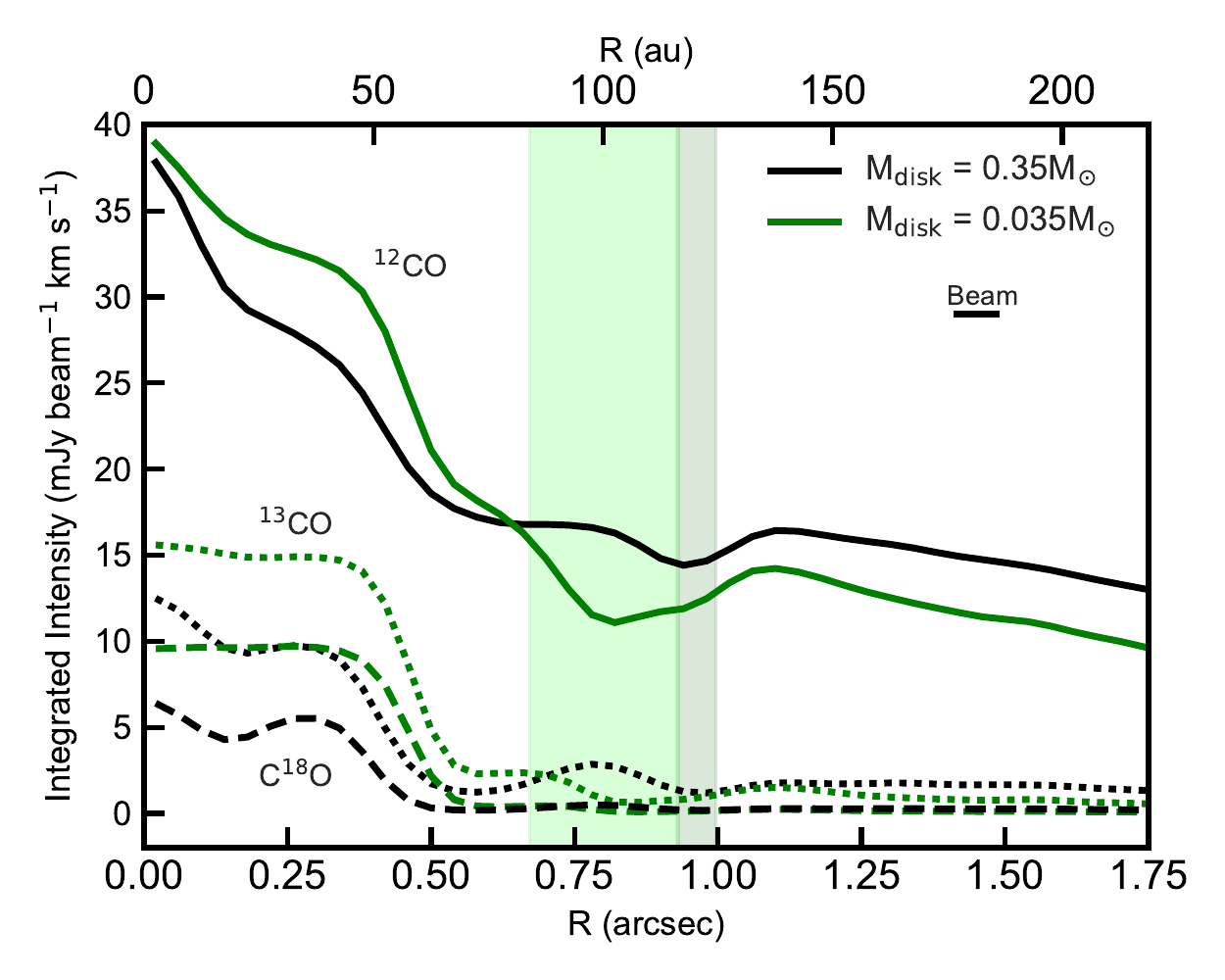}
\caption{We illustrate the difference for the J=2-1 transition of CO isotopologues with 0.1 arcsec resolution for two different disk masses. If we increase the disk mass by a factor of ten, the emission is decreased. The lines are optically thick for the massive disk, therefore we can infer that the temperature is actually higher in the less massive disk. Nevertheless, if the gap is deep enough, that trend can switch if one of the lines becomes optically thin, while the other stays optically thick, for example in $^{12}$CO.}
\label{Fig:Diskmasses}
\end{figure}

The comparison that is seen in Figure \ref{Fig:Diskmasses}  shows that for a very massive disk, the $^{12}$CO emission while showing slightly different structure has only a small change compared to the order of magnitude increase in the CO content.  
The radial profiles of $^{12}$CO emission show that the dip in the gap is shallower for the massive case. We associate the change in the trend to the fact that $^{12}$CO remains optically thick in the massive disk, while in the hotter less massive disk, the $^{12}$CO becomes optically thinner inside the gap, tracing deeper and colder layers. A similar behavior can be extrapolated to the other CO isotopologues, where the emission becomes comparable inside the gap. Therefore, the actual dip in the CO emission is related to the gas surface density  and the traced layer, where an excitation temperature effect can be observed as well. 

\subsubsection{Effect of Different Cosmic-ray Ionization Rates}\label{Sect:CR}

\begin{figure}
\includegraphics[width=.5\textwidth]{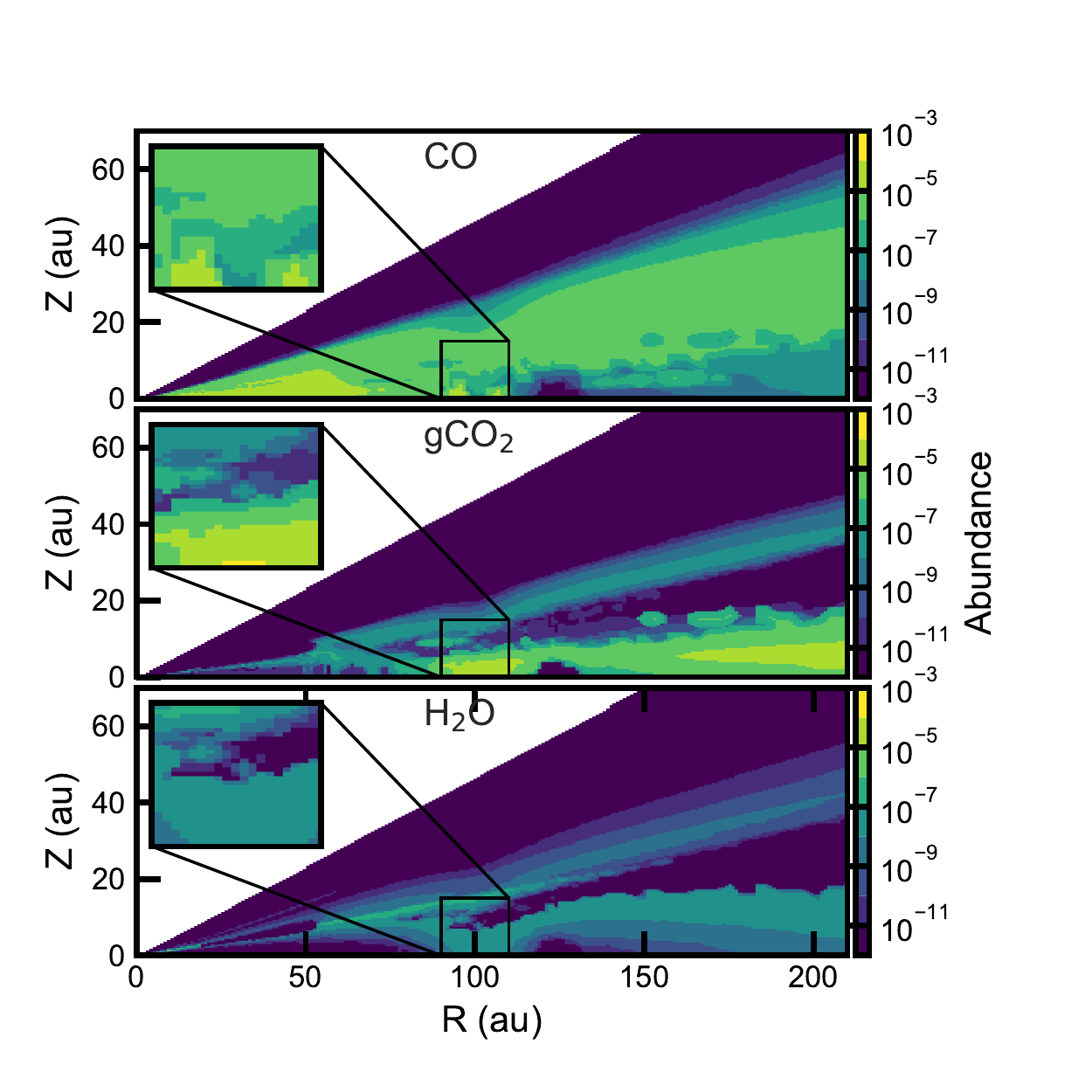}
\caption{The figure shows the abundance of gaseous CO and H$_2$O and frozen CO in models with the substructure, but without the presence of cosmic-rays in the disk. It shows that inside the gap and particularly in the midplane, CO is more abundant. It implies that cosmic-rays decrease the amount of CO in the midplane through induced photoreactions. We expect that cosmic-rays will also change the abundance for other molecules and carbon carriers, H$_2$CO and CH$_3$OH for instance \citep{De_barros_2014,Kalvans_2015}. }
\label{Fig:2d_noCR}
\end{figure}

Cosmic-rays are one of the main heating sources in the midplane, particularly at large radii. We explore the effect that cosmic-rays have in our models. Since cosmic-rays can reach deeper layers in the disk, they  cause the dissociation and desorption several species, they also ionize the medium enriching the reactions of photochemical tracers such as HCO$^{+}$ and H$_2$CO. We can see in Figure \ref{Fig:2d_noCR} that when cosmic-rays are removed, there is less CO desorption; however, it is still present, in particular at the deepest point of the gap. Even if CO and  H$_2$O are desorbed, they do not photodissociate in the midplane. On top of that, we expect that the presence of cosmic-rays will enhance the production of methanol (CH$_3$OH), becoming one of the main carbon carriers as the bulk chemistry evolves from small organics towards more complex molecules.

\section{Discussion}\label{sec:Disc and Con}

In general, our results suggest that the gap carved by the planet has a different thermal structure than in the undepleted case with a different radiation field as well. Given that the $\tau_{\rm UV}=1$ layer is pushed closer to the midplane (see Fig. \ref{Fig:sketch}), if the planet gets massive enough, UV photons could even reach the midplane. We  therefore expect more chemical interaction between the planet and the warm molecular layer, along with a higher ionization fraction around the planet. Such interaction would be enhanced with more massive planets. 

It is still a matter of discussion whether planets are the ones carving the gaps and creating the rings, or planets are being formed on those regions. However, the link between dust substructures and the changes in the physical and chemical structure of the disk they incite is a vital relationship to understand.

\subsection{Physical Structure}

The actual effect of the dust substructure will be dependent on the dynamical structure and properties of the host disk, and on the substructure itself. Nevertheless, our model is a reasonable starting point and allow us to qualitatively understand the general impact of a dust substructure on a protoplanetary disk.

\subsubsection{Location and Depth of Dust Substructure}

In our models we have explored the chemistry inside a gap-ring substructure at one location beyond the CO snowline in a young disk.  
If the gaps/rings are situated closer to the star, more volatiles would be present, enhancing photoprocessing in the gap, while within the ring, denser conditions would enhance any freeze-out. If the substructure were farther out in the disk, the ring would induce a less noticeable effect, but the gap could increase the sublimation of molecules through photodesorption. 

We predict that the depth or strength of the substructure is also important for the warm molecular layer shifting deeper into the disk. A deeper or more dust-depleted gap would increase the transparency of the gap to UV photons placing the warm molecular layer deeper as the small dust population is coupled to the gas. Such effect takes the gaseous molecules closer to the midplane. A central question for later exploration will be whether the gap is inside or outside the snowline, location of key volatile ices in the midplane, such as CO, CH$_4$, or CO$_2$.

\subsubsection{Meridional Flows and Mixing}\label{ssection: MF}

Several 3D hydrodynamical simulations show the presence of vertical flows inside the gaps carved by massive planets  \citep{Morbidelli_2014,Fung..Chiang..2016,Szulagyi..et..al}. These vertical flows, labeled meridional flows, move gas and small dust from the surface of the disk towards the midplane inside the gap and upwards inside the disk beyond gap edges. Meridional flows have been observationally confirmed by \cite{Teague_Bae_Bergin} and may therefore change the chemical composition of the gap as seen in our simulations.   The observational effects of such flows on the chemical properties of the gap is uncertain.  However, \cite{Szulagyi..et..al} show that planets accrete through inflows with significant vertical components. A strong vertical component in the accretion inflow  implies that the planet's composition could be enriched with the chemically active layer above and below  it, an effect that has also been explored by \cite{Cridland..et..al..2020}. A thorough description of meridional and radial flows in planet-forming regions is beyond the scope of this paper. However, we run 3D  simulations with the hydrodynamical code PLUTO \citep{PLUTO} to test and to obtain order-of-magnitude values for the mass flows in the substructure (see Appendix \ref{Appendix}).

Based on our simulations, we expect radial mass flows in the midplane going away from the planet to be of the order of $\sim$ 10-100 M$_{\oplus}$ Myr$^{-1}$. Therefore, there is some mixing between the gaseous CO in the warm molecular layer and the midplane in the gap via vertical flow, and from the midplane to the gap's edges via the radial component of the motion.  Thus, both dissociated photoproducts of gas and ice along with shielded molecules and ice-coated grains  will flow towards the midplane. 

\begin{figure}
\includegraphics[width=.48\textwidth]{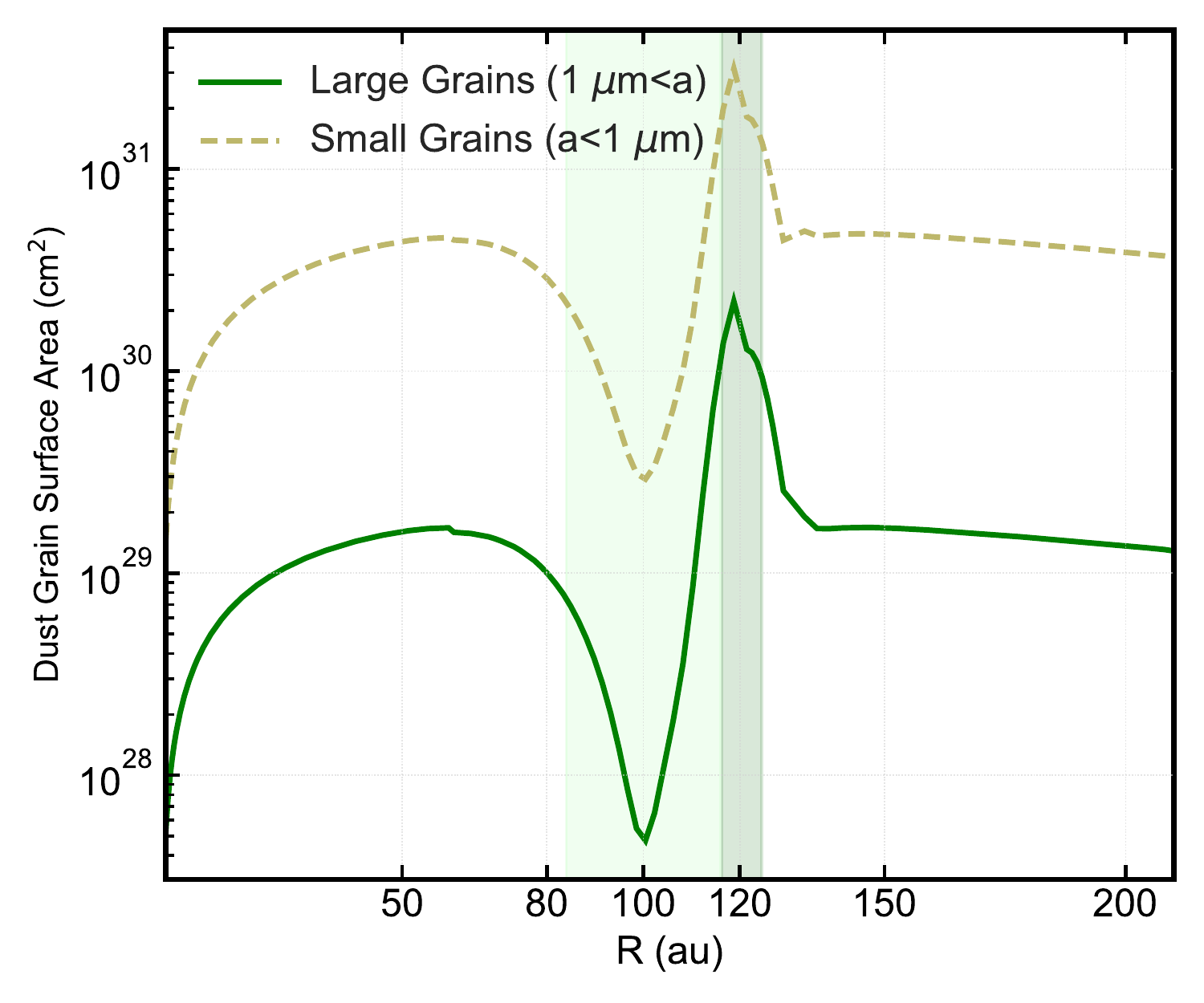}
\caption{The figure shows the available dust grain surface ($\int n_{a}\sigma_{a}dz$) area as a function of radius. We can observe in the figure  that dust grains in the ring increase the net surface area by a factor of ten. Therefore, it is likely that volatiles moving through radial decretion flows will probably be either adsorb, by the larger dust surface area in the ring, or freeze-out due to the low temperatures there. Even if large grains hold less surface area than small grains, given that they are concentrated in the midplane, they will have a bigger share of the surface area there, while small grains will be more important in the warm molecular layer.}
\label{Fig:Surf_Area}
\end{figure}

Further, the gaseous species that settle into the midplane but not onto the planet would be pushed towards the gap's edges (see \citealt{Kley..1999} for the velocity flow field inside the gap). The species drifting to the inner edge will either follow the gas flow returning to the midplane chemical equilibrium, or they will get trapped in an inner dust-rich ring. In the outer edge the volatiles may be photodissociated by the UV backscattering as they flow towards the outer pressure maxima, where the ring is present. Figure \ref{Fig:Surf_Area} shows how the surface area increases by two orders of magnitude between the gap and the ring. Throughout much of the disk, the surface area of solids is dominated by more numerous small grains; however, in the ring the probable dust-growth, dust settling, and therefore pile-up of large grains may lead the mm-sized particles to have a bigger share of the dust grains surface area, $n_{gr} \sigma$.  Thus, the ring, which is also locally colder, presents a freeze-out trap for this flow and all volatiles in the gas will likely be deposited on grain surfaces.  This will produce local enrichments of volatiles into ices depending on the overall temperature of the ring. However, whether this is a dominant effect is still unknown.  Among the species enriching the dust grains, we expect gaseous CO or C atoms to be present, which increases the C/O ratio within solids in the ring. This effect is important as the dust-rich rings represent likely sites for subsequent rounds of planetesimal formation and icy/rocky planet growth.

\subsection{Observational Footprints of the Dust Substructure Properties?}

Gas line emission is a central method to probe the physical properties within gaps and rings. Due to their abundance and subsequent emission brightness, CO isotopologues are main molecular tracers used to measure disk gas column density and temperature structure. However, chemical effects in planet-forming regions, i.e., inside dust substructures, change the relative abundance of CO to H$_2$ which lead to inaccurate measurements of the gas surface density. For example, $^{12}$CO emission is optically thick in general disk gas and is therefore a preferred temperature tracer \citep{Beckwith..et..al..1993,Schwarz..Bergin..2014,Weaver..et..al..2018}. In our simulations, the emission of $^{12}$CO coming from the gap is decreased by less than a factor of two (see Fig.\ref{Fig:syn_images}), but the actual gas surface density is decreased by an order of magnitude. Thus, the emission of $^{12}$CO keeps being optically thick and it is not a reliable tracer of the H$_2$ column density. This effect has been explored in more detail by \cite{Vandermarel..et..al..2019}. They show that the actual trend in the CO emission and its isotopologues has a strong dependence with the depth of the gap. 

Despite that $^{12}$CO by itself does not trace all the physical conditions in the disk, by using different CO isotopologues and other chemical tracers, we can discern the degeneracy between the chemistry and the physical conditions in the disk. Because CO isopologues trace different layers within the disk, observing them give us a first approach to the disk vertical structure. However, such approach is not enough given that they do not give the full chemical picture of the disk. The abundance of different molecular tracers is susceptible to local radiation fields and the relative abundances between volatile elements, such as C$_2$H or HCO$^+$ for example. To summarize, more than one molecular line is necessary to have a full thermo-chemical description around the planet forming regions in a protoplanetary disk.

\subsection{Comparison with Previous Models present in the Literature}\label{sect: Comparison}

There are previous works in the literature simulating the effect of dust substructures on the chemistry of  protoplanetary disks and the respective line emission predictions.  For instance, \cite{Facchini..et..al..2018} agree with some of our results in terms of the radiation field increasing inside the gap. In their simulations, they locate a planet carving a gap at 20 au including the 1D dust evolution code from \cite{Birnstiel_et_al_2010},which sets the dust size distribution and surface density in the substructure. They find that the $T_{\rm gas}/T_{\rm dust}$ ratio decreases in the gap, particularly in the midplane. Their result differs with ours, because their modeled dust-to-gas ratio inside the gap is lower. With such high dust depletions, the gas and the dust are not thermally coupled, leading to an increase in the dust temperature, while the gas is colder. Also, in our model we fixed the vertical structure to account for dust settling. Compared to \cite{Facchini..et..al..2018}, the dust-to-gas ratio in our models will be higher in the gap, and particularly in the midplane, leading to the differences in our thermal structure.  

\cite{Vandermarel..et..al..2018} study the CO emission for gaps created by snowlines or carved by planets. They illustrate the effect that different gap depths would have on the emission profiles of CO isotopologues. In terms of the physical properties, their models have assumed dust and gas depletions that encompass to our baseline model. In their modelling, the dust and gas temperature are still coupled in the denser layers of the disk, and they also predict a higher UV flux in the gap as well, although they do not have a ring in the models. They show that the CO emission inside the gap could increase or decrease depending on the depth of the substructure. Both \cite{Vandermarel..et..al..2018} and our model predict CO sublimation in gaps beyond the CO snow line. In order to probe the actual depth of the gap in the gas surface density, one must move towards a larger variety of molecular species to characterize such substructures. Thus, more information about the thermo-chemical and dynamical structure of the disk is needed.

A previous model by \cite{Favre..et..al..2019} focused on the same source has similar effects to the ones reported by us. A CO abundance enhancement inside the gaps is also observed, at least for the two most abundant CO isotopologues. Our simulations do not include isotope-selective photodissociation, which explains their C$^{18}$O abundance to decrease inside the gap. Their CO abundance agrees with our findings. Even considering the gas depletion inside the gap, the CO column density drop is not as significant as the gas surface density with a higher CO abundance.

\subsection{Limitations of our Simulations}\label{ssection: Limits}

Our thermo-chemical runs assume several simplifying assumptions. We only consider a static structure.  Thus, any dynamical evolution of the disk has not been taken into account. The main carriers of molecules through the disk will be $\sim$micron-sized dust grains and mm-sized pebbles. As pebbles drift, they would carry frozen molecules across the snow lines, changing the local composition and enriching the chemistry, mostly in the inner 10 au of a protoplanetary disk \citep{Booth_Ilee_2019}, or even further for hotter disks.

The static structure of our model does not include the drift of small grains onto the gap and the subsequent carbon grain destruction. The carbon grain destruction could be relevant if we consider the drifting of grains from the outer disk that could enrich the gaseous carbon chemistry inside the gap. \cite{Wei_et_al_2019} show that carbon grain destruction  would replenish the gas chemistry with carbon, while the oxygen gets frozen-out. However, they explore the region inside the inner 10 au. Nevertheless, carbon grain destruction can increase the C/O ratio generating a more carbon-driven chemistry, which may change the main carbon carriers inside the substructure and the CO abundance as well.

We only located the gap in one location, so its effect on the chemistry is dependent on the species present there. Further, the gap location with respect to the snowline location of key volatiles, CO or H$_2$O just to give a few examples, should have significant differences as the physical local conditions and chemical species present differ between snowlines. Such differences have not been explored in detail and would give us insights about the chemical differences between planets formed at different locations.

\section{Summary}\label{sec:Summ}

In this study we modeled the chemical evolution in a protoplanetary disk by adding a gap and a ring as dust substructure in the density profile of the disk. We found different effects on the disk physical and chemical conditions: 

\begin{enumerate}
    \item Where the disk is depleted in gas and dust, i.e., inside the gaps, it becomes more transparent to radiation, therefore heating up that region. At the same time,  high energy photons (UV and X-rays) penetrate deeper in the disk, so the photochemistry is enhanced at the gap.  One key effect is back-scattering off the outer edge of the gap.
    \item In the rings or dust-rich regions, the disk is shielded to radiation and high energy photons, and becomes relatively colder. The rings present a freeze-out trap for the molecules that would otherwise be in the gas phase, i.e, rings have the potential to lock-up some volatile species on the grains.
    \item Since the disk inside the gaps is more transparent, the UV photons and the cosmic-rays will  push the molecular layer closer to the midplane. Inside the gap, CO sublimation in the midplane from residual small grains that drift inside the gap increases the amount of gaseous CO in the locations of planet formation. Cosmic-rays, if present, enhance the production of CO$_2$ on the grains, rather than locking up the carbon atoms in other molecules. Molecular tracers of X-rays and UV radiation, such as HCO$^+$ and C$_2$H will also be produced, particularly in the middle of the gap, although we predict that it will depend on the gap's relative depth and the X-rays flux in the disk.
    \item  The dust grain surface area in the rings is larger by almost two orders of magnitude. Therefore, the gas-grain chemistry will be more important considering the freeze-out of gaseous molecules due to the lower temperature in the ring.
    \item The observational effect of the gap will be degenerated with its depth. Even if the local abundance of gaseous species is increased, their column density will be dependant on the local gas surface density. On top of that, most of the gas species are desorbed closer to the midplane. Given that the temperature in the midplane is usually lower than the one on the surface, the actual line emission may be lower, which could lead to a misinterpretation of gas depletion inside the gap. 
    \item The meridional flows in the disk are fast enough to carry the sublimated species in the gap towards its inner and outer edges. The meridional flow on top of radial flows in the midplane chemically enrich the dust grains and pebbles located at the edges of the gap. Such pebbles and dust grains are considered seeds for rocky cores that may end-up forming a planet.
    \item Although our simulation put the substructure in only one location, we expect its influence on the chemistry and the physical conditions to be significant. Nevertheless, its effect may vary depending on where it is located. 
\end{enumerate}

\software{\texttt{Astropy} \citep{astropy:2013, astropy:2018}, \texttt{rac2d}  \citep{Fujun..&..Bergin..2014}
}
%% Putting eqnarrays or equations inside the mathletters environment groups
%% the enclosed equations by letter. For instance, the eqnarray below, instead
%% of being numbered, say, (4) and (5), would be numbered (4a) and (4b).
%% LaTeX the paper and look at the output to see the results.

\acknowledgments

K.Z. acknowledges the support of NASA through Hubble Fellowship grant HST-HF2-51401.001 awarded by the Space Telescope Science Institute, which is operated by the Association of Universities for Research in Astronomy, Inc., for NASA, under contract NAS5-26555.     EAB is grateful to support for this work from the National Science Foundation via AST \#1907653


\vspace{5mm}
%\facilities{}




\appendix

\section{Hydrodynamical Tests}\label{Appendix}

 In order to get approximated mass flows values in the dust substructure we run hydrodynamical simulations with PLUTO \citep{PLUTO}. We consider a 0.09 Jupiter mass planet at 100 au and  an $\alpha$-viscosity, $\alpha$=10$^{-4}$, following a similar setup to the used by \cite{Zhang..et..al..2018} to fit the dust continuum emission of AS 209, although with a higher viscosity. The simulation runs for 1000 orbits, equivalent to 1 Myr with a planet at 100 au. After the 1000 orbits, the mass flows were calculated. As result of the vertical shear of the disk setup, the isothermal equation of state and low $\alpha$-viscosity set in our simulations, there is a significant influence from the vertical shear instability \citep{Nelson..et..al..2013} in the mass circulation of the disk.
 
 \begin{figure}
 \centering
\includegraphics[width=.65\textwidth]{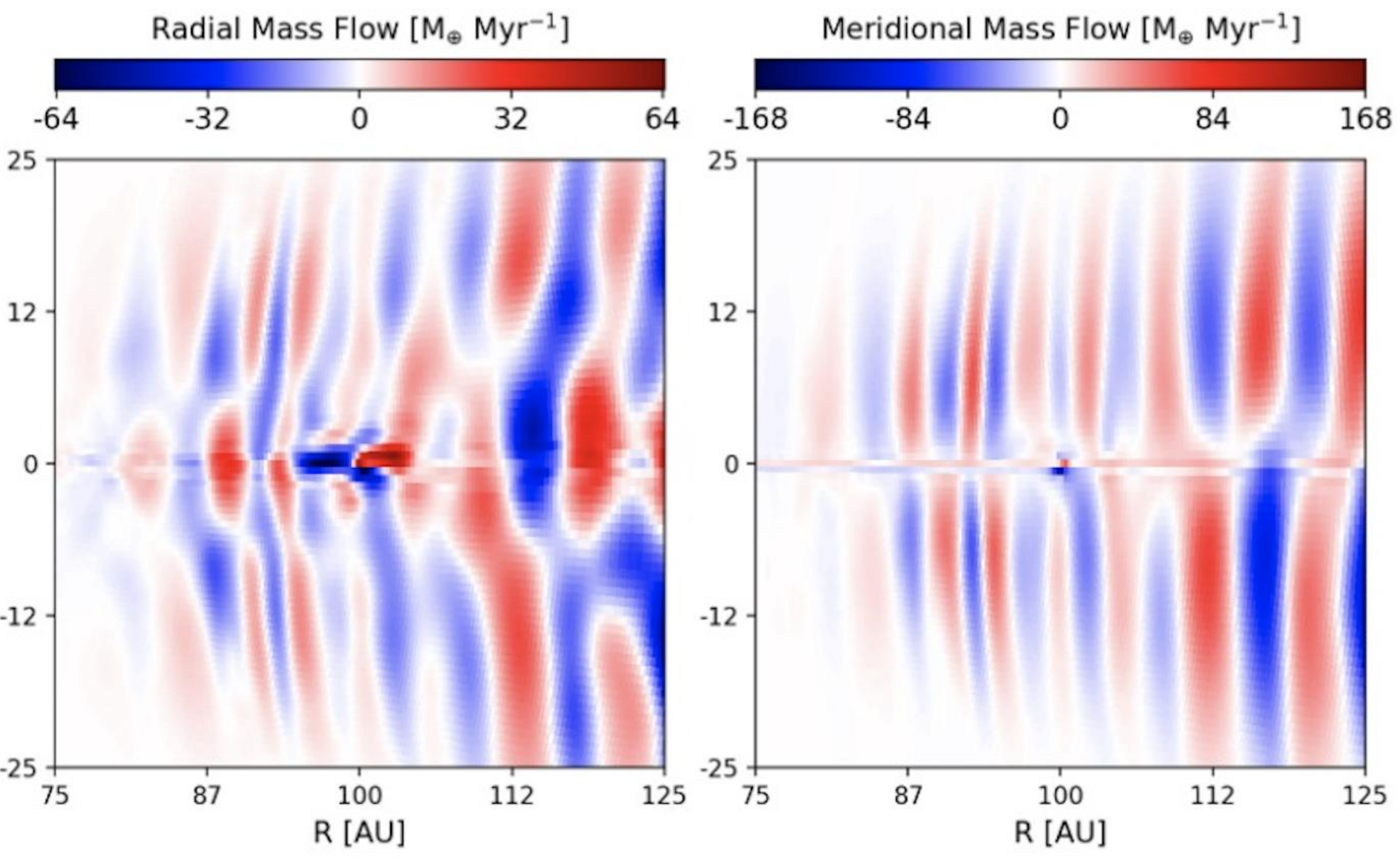}
\caption{We show the radial and meridional mass flows azimuthally-integrated for a 3D hydrodynamical simulation with a 0.09 Jupiter mass planet embedded at 100 au and $\alpha$=10$^{-4}$. Even for a small giant planet, the circulation around the planet's location in the midplane reaches values higher than 10 M$_{\oplus}$ Myr${-1}$ in the radial direction. The meridional mass flow is also large enough to exchange the chemical species from upper layers in the disk atmosphere with the midplane of the disk. The gaseous species relatively more abundant in the gap can get carried towards the gap edges and  dust-rich zones enriching dust grains and future rocky seeds.}
\label{Fig:hydro}
\end{figure}

In Figure \ref{Fig:hydro} we show the azimuthally-integrated radial and meridional flows, $(\int \rho v_r r^2d\phi) d\theta $ and $(\int \rho v_{\theta} rd\phi)dr $ respectively. We observe that the radial flow of mass in the midplane at 100 au is $\sim$ 50 M$_{\oplus}$ Myr$^{-1}$ . With such mass flow rate, in a thousand years, the CO photodissociation timescale in the gap, a mass of  0.05 M$_{\oplus}$ would move out the gap.  Therefore, the radial circulation  in and out the gap allows to chemically enrich the dust grains and dust grains located at its edges. The chemical species with large timescales ($\tau>10^5$ yr) survive long enough to get to change the composition of future rocky cores. For example, in our models, the gaseous CO abundance in the midplane at the center of the gap is of the order of 10$^{-6}$. If the mass flow if of the order of 10 M$_{\oplus}$ Myr$^{-1}$, the gaseous CO mass flow is 10$^{-5}$ M$_{\oplus}$ Myr$^{-1}$, which could be higher inside the CO snow line. When the CO reaches the border of the gaps to colder regions, it can be adsorbed by dust grains or freeze-out in cold regions of the midplane.

Besides radial flows, there is also meridional circulation. The vertical  velocity in the middle of the gap would be of the order of 0.1c$_s$, with c$_s$ the local sound speed \citep{Fung..Chiang..2016}. Assuming that velocity, the meridional flow timescale  is $\tau_{\rm mf}$ = H/0.1c$_s = 10/\Omega \approx \tau_{\rm orbit}$ . At 100 au, the orbital timescale for our stellar parameters is $\tau_{\rm orbit} \approx 1000$ yr. Under a timescale of 1000 yr, the species at different substructures of the disk can be carried from one to the other, effectively enriching regions where otherwise they would not be present.

 It is noteworthy that the parameters of the hydrodynamical simulation used a low viscosity value with a planet mass of roughly 0.09 Jupiter masses. A more massive planet and a higher viscosity would only increase the mass flow and shorten the timescales for the  circulation. With shorter circulation timescales, the chemical equilibrium and chemical enrichment will probably differ.




\bibliographystyle{aasjournal}
\bibliography{ref}



\end{document}